\documentclass[numberedappendix]{emulateapj}
\usepackage{amsmath}
\usepackage{rotating}

\usepackage{epsfig}
\newcommand{\saxj}{\mbox{SAX J1808.4$-$3658}}
\newcommand{\RXTE}{\textit{RXTE}}
\newcommand{\us}{$\mu$s}
\newcommand{\uHz}{$\mu$Hz}
\newcommand{\fluxunits}{~erg~cm$^{-2}$~s$^{-1}$}
\newcommand{\tempo}{\textsc{tempo}}
\newcommand{\Tempo}{\textsc{Tempo}}
\newcommand{\pz}{\phantom{0}}

\begin{document}

\shortauthors{Hartman et~al.}
\shorttitle{Timing of \saxj}

\submitted{Accepted for publication in ApJ}
\title{The long-term evolution of the spin, pulse shape, and orbit \\
of the accretion-powered millisecond pulsar SAX~J1808.4$-$3658}
\author{Jacob M. Hartman\altaffilmark{1}, Alessandro Patruno\altaffilmark{2},
Deepto Chakrabarty\altaffilmark{1}, David L. Kaplan\altaffilmark{1,6}, Craig
B. Markwardt\altaffilmark{3,4}, Edward H. Morgan\altaffilmark{1}, Paul
S. Ray\altaffilmark{5}, Michiel van der Klis\altaffilmark{2}, and Rudy
Wijnands\altaffilmark{2}}
\altaffiltext{1}{Department of Physics and Kavli Institute for Astrophysics
and Space Research, Massachusetts Institute of Technology, Cambridge, MA
02139; jhartman,deepto,ehm@space.mit.edu}
\altaffiltext{2}{Astronomical Institute ``Anton Pannekoek'' and Center for
High Energy Astrophysics, University of Amsterdam, Kruislaan 403, 1098 SJ
Amsterdam, Netherlands; apatruno,michiel, rudy@science.uva.nl}
\altaffiltext{3}{CRESST/Department of Astronomy, University of Maryland, 
College Park, MD 20742}
\altaffiltext{4}{Astrophysics Science Division, NASA Goddard Space Flight 
Center, Greenbelt, MD 20771; craigm@milkyway.gsfc.nasa.gov}
\altaffiltext{5}{Space Science Division, Naval Research Laboratory, 
Washington, DC 20375, USA; paul.ray@nrl.navy.mil}
\altaffiltext{6}{Pappalardo Fellow}

\begin{abstract}

We present a 7~yr timing study of the 2.5~ms X-ray pulsar \saxj, an X-ray
transient with a recurrence time of $\approx$2~yr, using data from the {\em
Rossi X-ray Timing Explorer} covering 4 transient outbursts (1998--2005).  We
verify that the 401~Hz pulsation traces the spin frequency fundamental and not
a harmonic.  Substantial pulse shape variability, both stochastic and
systematic, was observed during each outburst.  Analysis of the systematic
pulse shape changes suggests that, as an outburst dims, the X-ray ``hot spot''
on the pulsar surface drifts longitudinally and a second hot spot may appear.
The overall pulse shape variability limits the ability to measure spin
frequency evolution within a given X-ray outburst (and calls previous
$\dot\nu$ measurements of this source into question), with typical upper
limits of $|\dot\nu| \lesssim 2.5\times10^{-14}$~Hz~s$^{-1}$ ($2\,\sigma$).
However, combining data from all the outbursts shows with high ($6\,\sigma$)
significance that the pulsar is undergoing long-term spin down at a rate
$\dot{\nu}=(-5.6\pm2.0)\times10^{-16}$~Hz~s$^{-1}$, with most of the spin
evolution occurring during X-ray quiescence.  We discuss the possible
contributions of magnetic propeller torques, magnetic dipole radiation, and
gravitational radiation to the measured spin down, setting an upper limit of
$B<1.5\times10^8$~G for the pulsar's surface dipole magnetic field and
$Q/I<5\times10^{-9}$ for the fractional mass quadrupole moment.  We also
measured an orbital period derivative of $\dot{P}_{\rm orb} =
(3.5\pm0.2)\times10^{-12}$~s~s$^{-1}$.  This surprising large $\dot{P}_{\rm
orb}$ is reminiscent of the large and quasi-cyclic orbital period variation
observed in the so-called ``black widow'' millisecond radio pulsars.  This
further strengthens previous speculation that \saxj\ may turn on as a radio
pulsar during quiescence.  In an appendix we derive an improved (0.15~arcsec)
source position from optical data.
\end{abstract}

\section{Introduction}

The growing class of accretion-powered millisecond X-ray pulsars discovered by
the {\em Rossi X-Ray Timing Explorer} (\RXTE) has verified the hypothesis that
old millisecond pulsars obtained their rapid spins through sustained accretion
in X-ray binaries.  These objects provide a versatile laboratory.  The X-ray
pulse shapes arising from the magnetically channeled accretion flow can
constrain the compactness (and hence the equation of state) of the neutron
star.  Tracking the arrival times of these X-ray pulses allows us to measure
the pulsar spin evolution, which directly probes magnetic disk accretion
torque theory in a particularly interesting regime \citep{Psaltis99} and also
allows exploration of torques arising from other mechanisms such as
gravitational wave emission \citep{Bildsten98}.  There have been several
reports of significant spin evolution in accreting millisecond pulsars, some
with implied torques that are difficult to reconcile with standard accretion
torque theory \citep{Markwardt03, Morgan03, Burderi06, Burderi07}.  However, a
variety of effects (including limited data spans, pulse shape variability, and
non-Gaussian noise sources) can complicate the interpretation of these
measurements.  In this paper, we address these difficulties using a
comprehensive analysis of the most extensive data set.

Of the eight accretion-powered millisecond pulsars currently known, the first
one remains the best-studied example. The X-ray transient \saxj\ was
discovered during an outburst in 1996 by the {\em BeppoSAX} Wide Field Cameras
\citep{IntZand98}.  Timing analysis of \RXTE\ data from a second outburst in
1998 revealed the presence of a 401~Hz (2.5~ms) accreting pulsar in a 2~hr
binary \citep{Wijnands98, Chakrabarty98}.  The source is a recurrent X-ray
transient, with subsequent $\approx$1 month long X-ray outbursts detected in
2000, 2002, and 2005; it is the only known accreting millisecond pulsar for
which pulsations have been detected during multiple outbursts.  Faint
quiescent X-ray emission has also been observed between outbursts, although no
pulsations were detected \citep{Stella00, Campana02, Heinke07}.  A source
distance of 3.4--3.6~kpc is estimated from X-ray burst properties
\citep{IntZand01, Galloway06}.  The pulsar is a weakly magnetized neutron star
($[1$--$10]\times10^8$~G at the surface; \citealt{Psaltis99}) while the mass
donor is likely an extremely low-mass ($\approx$0.05~$M_\sun$) brown dwarf
\citep{Bildsten01}.  \saxj\ is the only source known to exhibit all three of
the rapid X-ray variability phenomena associated with neutron stars in LMXBs:
accretion-powered millisecond pulsations \citep{Wijnands98}, millisecond
oscillations during thermonuclear X-ray bursts \citep{Chakrabarty03}, and
kilohertz quasi-periodic oscillations \citep{Wijnands03}.  After submitting
this paper, we learned of an independent analysis by \citet{DiSalvo07b}
reporting an increasing orbital period (see \S\ref{sect:porbderiv}).

An optical counterpart has been detected both during outburst \citep{Roche98,
Giles99} and quiescence \citep{Homer01}.  The relatively high optical
luminosity during X-ray quiescence has led to speculation that the neutron
star may be an active radio pulsar during these intervals \citep{Burderi03,
Campana04}, although radio pulsations have not been detected \citep{Burgay03}.
Transient unpulsed radio emission \citep{Gaensler99, Rupen05} and an infrared
excess \citep{Wang01, Greenhill06}, both attributed to synchrotron radiation
in an outflow, have been reported during X-ray outbursts.

In this paper, we describe our application of phase-connected timing solutions
for each outburst to study the spin history and pulse profile variability of
\saxj, providing the first look at the evolution of an accretion-powered X-ray
pulsar.  In section~\ref{sect:analysis}, we outline our analysis methods,
noting the difficulties raised by pulse profile noise and describing a new
technique to obtain a minimum-variance estimate of the spin phase in the
presence of such noise.  In section~\ref{sect:results}, we present the results
of this analysis.  In particular, we observe that the source is spinning down
between outbursts, the binary orbital period is increasing, and the pulse
profiles change in a characteristic manner as the outbursts progress.
Finally, in section~\ref{sect:discussion}, we discuss the implications of
these results to the properties of the neutron star and accretion geometry.

\section{X-ray Observations and Data Analysis}
\label{sect:analysis}

\subsection{\RXTE\ data reduction}

The \RXTE\ Proportional Counter Array \citep[PCA;][]{Jahoda96} has repeatedly
observed \saxj, primarily during outburst.  These observations total 307
separate pointings and an exposure time of 1,371 ks from 1998 through 2005.
The PCA comprises five identical gas-filled proportional counter units (PCUs)
sensitive to X-rays between 2.5 and 60 keV.  Each PCU has an effective area of
1200 cm$^2$.  It is uncommon for all five PCUs to be active: some are
periodically disabled to decrease their rates of electrical breakdown
\citep{Jahoda06}.  The average number of active PCUs has declined as the
\RXTE\ ages, and most observations during the 2002 and 2005 outbursts of
\saxj\ only include two or three.

All but three of the observations of \saxj\ were taken with the
\verb!E_125US_64M_0_1S! mode, which records the arrival of each photon with a
time resolution of 122~\us\ and 64 energy channels covering the full range of
the detectors.  The other observations were rebinned to be compatible with the
122~\us\ resolution data; using higher time resolutions provides no benefit.
We shifted the photon arrival times to the Earth's geocenter\footnote{We
shifted the photons to the geocenter rather than the solar system barycenter
since the \tempo\ pulse timing program, which we used to fit phase models to
the arrival times, was designed for radio timing and thus expects photon
arrival times at some point on the Earth.  \Tempo\ itself performed the
barycentric corrections using the quoted position.} using our improved optical
position of R.A.~= $18^{\rm h}08^{\rm m}27\fs62$, Decl.~=
$-36\degr58\arcmin43\farcs3$ (equinox J2000), with an uncertainty of
$0\farcs15$.  (Please refer to Appendix~\ref{app:newpos} for details on this
improved position.)  We then applied the \RXTE\ fine clock correction, which
provides absolute time measurements with errors of less than 3.4~\us\
\citep[99\% confidence;][]{Jahoda06}.  Finally, we filtered the data to remove
Earth occultations, intervals of unstable pointing, and thermonuclear X-ray
bursts.  For three observations at the start and end of the 1998 outburst, we
relaxed our requirement of stable pointing and included raster scanning data
to extend our baseline for measuring the frequency evolution during this
outburst.  These observations provided additional valid phase measurements,
but they were not used to calculate fractional amplitudes since the
contribution of the source and background varied as the \RXTE\ panned across
the source.  Table~\ref{tbl:obslist} lists all the observations that we
included in our analysis.

\begin{deluxetable*}{lcrrcl}
\tabletypesize{\footnotesize}
\tablecolumns{6}
\tablewidth{0pt}
\tablecaption{Observations analyzed for each outburst\label{tbl:obslist}}
\tablehead{
  \colhead{} &
  \colhead{Data range} &
  \colhead{\#} &
  \colhead{Time} &
  \colhead{Avg. \#} &
  \colhead{Observation IDs}\\
  \colhead{} &
  \colhead{(MJD)} &
  \colhead{obs.} &
  \colhead{(ks)} &
  \colhead{PCUs} &
  \colhead{}
}
\startdata
1998 Apr & 50914.8 -- 50939.6 &  21 & 178.1 & 4.67 & {\tt 30411-01-*}\\
2000 Feb & 51564.1 -- 51601.9 &  38 & 126.8 & 3.74 & {\tt 40035-01-01-00} -- {\tt 40035-01-04-01}\\
         &                    &     &       &      & {\tt 40035-05-02-00} -- {\tt 40035-05-18-00}\\
2002 Oct & 52562.1 -- 52602.8 & 129 & 714.5 & 3.25 & {\tt 70080-01-*}, {\tt 70080-02-*}\\
         &                    &     &       &      & {\tt 70080-03-05-00} -- {\tt 70080-03-24-00}\\
         &                    &     &       &      & {\tt 70080-03-25-01}, {\tt 70518-01-*}\\
2005 Jun & 53523.0 -- 53581.4 &  55 & 284.3 & 2.84 & {\tt 91056-01-01-01} -- {\tt 91056-01-04-01}\\
         &                    &     &       &      & {\tt 91418-01-01-00} -- {\tt 91418-01-07-00}\smallskip
\enddata
\tablecomments{The ranges of observation IDs given here are for numerically
sorted IDs, which do not always reflect temporal order.}
\end{deluxetable*}

We consistently used an energy cut of roughly 2--15~keV for our timing
analysis.  While the source is readily detectable in the PCA at higher
energies, the background dominates above 15~keV, especially in the dimmer
tails of the outbursts.  Excluding these high-energy counts optimized the
detection of pulsations when the source was dim, providing a longer baseline
for our timing analysis.

\subsection{Pulse timing analysis with \tempo}

The core of this analysis resembles the work long-done for radio pulsars and
slowly rotating X-ray pulsars.  We first folded intervals of data according to
a phase timing model to obtain pulse profiles.  We next compared the profile
from each interval to a template profile in order to calculate the offset
between the observed and the predicted pulse times of arrival (TOAs).  We then
improved the initial phase model by fitting it to these TOA residuals.

We used the \tempo\ pulsar timing
program\footnote{\url{http://www.atnf.csiro.au/research/pulsar/tempo/}},
version 11.005, to calculate pulse arrival times from a phase model and to
improve a phase model by fitting it to arrival time residuals.  \Tempo\ reads
in a list of TOAs and a set of parameters describing the pulsar timing model.
It then adjusts the model to minimize the timing residuals between the
predicted and observed arrival times.  The output files include a revised
timing model, a covariance matrix for the fit parameters, and a list of the
timing residuals.  \Tempo\ also includes a predictive mode, which takes a
timing model and generates a series of polynomial expansions that give the
model's pulse arrival times during a specified time interval.  \Tempo\ has
been a standard tool of the radio pulsar community for decades and is
well-tested at the microsecond-level accuracies with which we are measuring
TOAs.

Our timing models fit for the following parameters: the pulsar spin frequency
and (if necessary) the first-order frequency derivative; the times and
magnitudes of any instantaneous changes in the frequency; and the orbital
parameters.  Our models supplied, but did not fit, the position of the source
from Appendix~\ref{app:newpos}.  Because we fit the outbursts separately, the
$\approx$1~month of data that each provides was not sufficient to improve the
source position: the position of \saxj\ (in particular, its right ascension)
is degenerate with the frequency and frequency derivative on such timescales.

To parametrize the orbit of \saxj, we used \tempo's {\tt ELL1} binary model,
which employs the Laplace parameters $e \sin \omega$ and $e \cos \omega$,
where $e$ is the eccentricity and $\omega$ the longitude of periastron
passage.  This parametrization avoids the degeneracy of $\omega$ in
low-eccentricity systems \citep{Deeter81}.  For most of the fits, we held $e =
0$ and solely fit the projected semimajor axis $a_{\rm x}\sin i$, the orbital
period $P_{\rm orb}$, and the time of ascending node\footnote{Past pulsar
timing of \saxj\ uses the $T_{90}$ fiducial, marking a time at which the mean
longitude is 90\degr.  Since its orbit is circular, $T_{90} = T_{\rm asc} +
P_{\rm orb} / 4$.} $T_{\rm asc}$.  As a test of this assumption, we also
repeated the fits allowing $e$ to vary.  It was always consistent with zero.

This analysis depends critically on the accurate calculation and processing of
the TOAs.  To verify our results, they were independently calculated using two
entirely separate data pathways.  One corrected the \RXTE\ count data to the
geocenter and used \tempo\ to barycenter the TOAs, as described in the
previous section; the other used the
FTOOL\footnote{\url{http://heasarc.gsfc.nasa.gov/ftools/}} {\tt faxbary} to
barycenter the count data.  Independent codes were then used to divide the
count data into 512~s intervals, fold it according to a phase model, and
measure the pulse times of arrival.  Finally, we used both \tempo\ and its
replacement, {\sc tempo2} \citep{Tempo2}, to process the TOAs and refine the
timing models.  In all cases, the agreement between the final timing models
was good.

\subsection{TOA calculation in the presence of profile noise}
\label{sect:toacalc}

Special care must be taken when measuring the pulse TOAs for rapidly rotating
accretion-powered pulsars.  In these systems, the pulse profiles exhibit
variability on timescales of $\sim$10~hr and longer that is well in excess of
the Poisson noise expected from counting statistics.  In this section, we
develop a procedure to obtain a minimum-variance estimate of the timing
residuals in the presence of such noise.

We use the term ``noise'' with respect to spin timing analysis simply to mean
phase variability of one or more harmonics that does not seem to be due to
underlying spin frequency changes.  While some of this profile variability may
in fact be quite ordered --- distinctive pulse shape changes that occur in
every outburst, for instance --- we cannot model all of them and thus consider
the unmodeled profile variability as ``noise'' from the phase-timing
perspective.  In this section, we attempt to minimize the impact of such
variability on the accuracy of our pulse arrival times by favorably weighting
data from less-noisy harmonics; in \S\ref{sect:paramfitting} we describe a
Monte Carlo technique to estimate its impact on the timing model parameters.

To calculate the TOAs, we divided the timing data into 512~s intervals and
determined one pulse arrival time per interval.  We chose this length because
it provides sufficient counts to make accurate measurements in the dim tails
of the outbursts, while it still is short enough to sample within the 7249~s
orbital period.  This is necessary to improve the binary model and resolve any
additional short-timescale variability.

For each outburst, we used \tempo's predictive mode to generate a series of
polynomial expansions predicting the times of pulse arrivals at the geocenter.
These ephemerides are based on the revised optical position, our best-known
orbital parameters, and a simple, constant-frequency spin model.  Using the
expansions, we calculated the expected phase for each photon arrival time.
For each 512~s interval, we then divided the phases into $n$ phase bins in
order to create folded pulse profiles.

We then decomposed the profiles into their Fourier components.  For a given
folded profile, let $x_j$ designate the number of photons in the $j$th phase
bin, and $N_{\rm ph} = \sum_{j=1}^n x_j$ is the total number of photons.  The
complex amplitude of the $k$th harmonic is then
\begin{equation}
  a_k = \sum_{j=1}^n x_j \exp\left(2\pi i j k / n\right) \, .
\end{equation}
Throughout this paper, we number the harmonics such that the $k$th harmonic is
$k$ times the frequency of the 401~Hz fundamental.  Since our analysis for the
most part handles the phases and amplitudes of harmonics separately, we define
these quantities explicitly as follows:
\begin{equation}
  A_k \exp\left[2\pi i k \left(\phi_k + \Delta\phi_k\right)\right]
    = 2 a_k \, .
  \label{eq:folding}
\end{equation}
Here we are interested in the amplitude\footnote{We define $A_k$ such that it
is the actual amplitude, in photons, of the observed pulsations.  We must
therefore include both the positive and negative frequency components (which
are equal for real signals), introducing the factor of 2 on the right-hand
side of eqs.~(\ref{eq:folding}) and~(\ref{eq:templatedef}).} $A_k$ and the
phase residual $\Delta\phi_k$, which we measure relative to a fixed phase
offset $\phi_k$.  We include these offsets because we are principally
interested in measuring the phase deviations from a fixed template profile,
which we obtain by transforming the overall folded pulse profile from an
outburst:\footnote{In the presence of sudden pulse profile changes, we may use
multiple templates during a single outburst, thus using different values of
$\phi_k$ on either side of the change.  Further description is at the end of
this section.}
\begin{equation}
  A'_k \exp\left(2\pi i k \phi_k\right) =
    2 \sum_{j=1}^{n} x'_j \exp\left(2\pi i j k / n\right) \, .
    \label{eq:templatedef}
\end{equation}
Here $x'_j$ and $N'_{\rm ph}$ give the phase bin counts and total counts for
the template pulse profile.

Note that we define the phases such that shifting a fixed pulse profile by
some phase $\Delta\phi$ produces the same shift in the phase of each of its
harmonic: $\Delta\phi_k = \Delta\phi$.  Hence the unique phases for each
$\Delta\phi_k$ range from 0 to $1/k$.  Positive phase residuals corresponding
to time lags:  $\Delta\phi_k > 0$ indicates that the $k$th harmonic arrived
later than predicted by the model.

The uncertainty in the phase residuals $\Delta\phi_k$ due to Poisson noise
(derived in Appendix~\ref{app:phaseerrs}) are
\begin{equation}
  \sigma_k = \frac{\sqrt{2 N_{\rm ph}}}{2\pi k A_k} \, .  \label{eq:sigmak}
\end{equation}
For our analysis, we rejected phase measurements with uncertainties greater
than 0.1~ms (i.e., 0.04 cycles).  Generally, this cut only removed points in
the tails of the outbursts, where the flux was low.

The measured fractional rms amplitudes are
\begin{equation}
  r_k = \frac{A_k}{\sqrt{2}\left(N_{\rm ph} - B\right)} \, ,
\end{equation}
where $B$ is the approximate number of background events within our energy
range and time interval, estimated using the FTOOL {\tt pcabackest}.\footnote{
\url{http://heasarc.gsfc.nasa.gov/docs/xte/recipes/pcabackest.html}} The $r_k$
add in quadrature: the total rms fractional amplitude for a pulse profile
described with $m$ harmonics is $r = \left(\sum_{k=1}^m r_k^2\right)$$^{1/2}$.
Uncertainties on the fractional amplitudes are computed using the method
described by \citet{Groth75} and \cite{Vaughan94}, which accounts for the
addition of noise to the complex amplitude of the signal.  The probability
that the detection of a harmonic is due solely to Poisson noise is
$\exp(-P_k)$, where $P_k = \frac{1}{4} A_k^2 / N_{\rm ph}$ is the
unit-normalized power for the $k$th harmonic.  For a fuller review of Fourier
techniques in X-ray timing, we defer to \citet{VanDerKlis89}.

Since each harmonic provides an independent measurement of the phase residual,
we can combine them to provide the overall phase residual for the sample
pulse.  We obtain the optimal estimator by weighting each measurement
according to its variance:
\begin{eqnarray}
  \Delta\phi
      &=& \sum_{k=1}^m w_k\Delta\phi_k \ \Big/\ \sum_{k=1}^m w_k \, ; \\
  w_k &=& \frac{1}{\sigma_k^2} = \frac{k^2 A_k^2}{N_{\rm ph}} \, .
    \label{eq:simpleweights}
\end{eqnarray}
Thus far, this analytical method closely parallels the work long done on
spin-powered pulsars \citep[e.g.,][Appendix~A]{Taylor93}.

However, there are some essential differences that must be taken into account
when dealing with accretion-powered pulsars.  Unlike spin-powered pulsars,
which usually show one or more sharp, asymmetric pulses per cycle, there is
little harmonic content in the pulsations of \saxj\ beyond $k = 2$, so we
truncate the series there.  Additionally, while the individual pulses of radio
pulsars show appreciable variability from one period to the next, their
integrated profiles are very stable \citep{Manchester77}.  In such cases, one
expects that the pulse fractions of sample pulses are similar to the template
($A_k/N_{\rm ph} \approx A'_k/N'_{\rm ph}$) and that the harmonics reflect a
common phase residual ($\Delta\phi_k \approx \Delta\phi$) that traces the
rotational phase of the star.  Indeed, the standard template-matching analysis
is predicated on these assumptions.  Furthermore, any variability is assumed
to be due to Poisson noise, which is of equal magnitude at all timescales
(i.e., it is white noise).  In contrast, the accretion-powered pulsars show
substantial pulse profile variability.  Beyond the usual Poisson noise
($\sigma_k$ from eq.~[\ref{eq:sigmak}]), three additional issues complicate
the usual approach of template matching:  long-timescale correlations (i.e.,
red noise) in the observed pulse fractions, with each harmonic's $A_k$ varying
independently; red noise in the phase offsets $\Delta\phi_k$; and sudden pulse
profile changes, in which the phase offset between the two measured harmonics
changes drastically on the timescale of the observations.

In their timing analysis of 283~s accretion-powered pulsar Vela~X-1,
\citet{Boynton84} partially address the issue of intrinsic pulse profile noise
in the harmonics.  In the most general case, the amplitude of the variability
in the phase residuals $\Delta\phi_k$ is different for each harmonic $k$.
This is the case for both Vela~X-1 and \saxj.  They correct for the harmonic
dependence of these fluctuations by scaling the phase residuals of each
harmonic by constants chosen such that the phase residuals all have the same
amplitude of variability \citep{Boynton85b}.  Thus the influence of
particularly noisy harmonics was diminished, and they were able to measure
with much greater accuracy the underlying spin of Vela~X-1.

Our approach was similar.  For each outburst, we measured the total rms
amplitude of the phase residuals for each harmonic with respect to a best-fit
constant-frequency model.  These residuals will represent the combined effect
of the Poisson noise and any intrinsic profile noise:
\begin{equation}
  \sigma^2_{k,{\rm rms}} = \left<\sigma^2_k\right>
                         + \sigma^2_{k,{\rm int}} \, .
  \label{eq:noisesigmas}
\end{equation}
We calculated the Poisson contribution $\left<\sigma^2_k\right>$ as a weighted
mean\footnote{To calculate the mean of the variances $\sigma^2_k$, we weight
each according to equation~(\ref{eq:simpleweights}).  Labeling each
uncertainty as $\sigma_{kl}$ for intervals $l = 1, \dots, N_{\rm int}$, we
have $\left<\sigma^2_k\right> = \left(\sum_{l=1}^{N_{\rm int}}
\sigma_{kl}^{-2} \big/ N_{\rm int}\right)^{-1}$.  This weighting scheme
prevents large variances during the tails of the outbursts from skewing the
results.} of the results from equation~(\ref{eq:sigmak}), giving us a value
for $\sigma^2_{k,{\rm int}}$.  We then incorporate this additional uncertainty
into our weighting to determine $\Delta\phi$:
\begin{equation}
  w_k = \frac{1}{\sigma_k^2 + \sigma^2_{k,{\rm int}}} \,.
  \label{eq:weighting}
\end{equation}
$\sigma_k^2$ changes from one TOA measurement to the next due to the
variability of the pulse fraction and count rate; there is no assumption that
these are constant, as there is in the case of standard template fitting.
$\sigma^2_{k,{\rm int}}$ is a constant measured independently for each
harmonic of each outburst.  The result is a minimum-variance estimator for
$\Delta\phi_k$.  For instance, if the 802 Hz second harmonic has smaller
intrinsic fluctuations than the fundamental, then our method presumes that it
better reflects the spin of the NS and will weight it more strongly.

Sudden pulse profile changes are somewhat simpler to deal with.  We use a
different template pulse profile (and hence different measurements on either
side of the change of $\phi_k$, defined in eq.~[\ref{eq:templatedef}])\@.  We
only modeled one such sudden profile change in this way: at the end of the
main body of the 2002 outburst (around MJD 52576), the fundamental phase
$\phi_1$ experienced a shift while the second harmonic, $\phi_2$, remained
constant.  The stability of $\phi_2$ allowed us to phase connect across the
feature, as \citet{Burderi06} also noted.

Our distinction between sudden profile changes and pulse profile noise is
admittedly somewhat arbitrary.  We make it solely in the interest of best
estimating the rotational phase of the star --- we are not claiming to model
some underlying difference in physical processes.  In 2002, the phase
residuals on either side of the modeled pulse profile change were quite
stable, albeit with different values of $\phi_1$.  This stability in both
harmonics makes it a good candidate for such treatment.  In contrast, the
phase residuals of both harmonics during the 2005 outburst show greater
amplitude fluctuations at nearly all timescales.  While its residuals and the
2002 residuals follow a similar pattern at the end of the main body of the
outburst (the phases of the second harmonic remain roughly constant, while the
phases of the fundamental drop appreciably), the phase of the fundamental
continues to fluctuate wildly after this event rather than settling down on a
``new'' template profile.  Therefore we elect to attribute these profile
changes to intrinsic noise and weight the relatively stable second harmonic
more strongly.

However, when the phases of {\em both} harmonics are continuously changing,
the ability to define pulse arrival times breaks down, and the data are of
little use for determining the spin of the star.  For \saxj, the pulse profile
is changing throughout the rises and peaks of the outbursts, so we excluded
these data from our measurements of the spin frequency.  We did include these
data when calculating the orbital parameters.  Since the timescale for these
pulse profile changes ($\gtrsim 10$~hr) is many times the orbital period, they
tend to average out and have little impact on these measurements.

The resulting phase residuals give the best estimator for the offset between
the measured and predicted pulse arrival times.  By adding these offsets to
the phases predicted by the \tempo\ ephemerides, we arrived at more accurate
pulse arrival times for each interval.

\subsection{Parameter fitting and uncertainty estimation}
\label{sect:paramfitting}

After measuring the pulse times of arrival, we input them into \tempo\ to
refit the timing solution.  In order to interpret the resulting models, we
must understand the nature of the noise in the TOAs and how it affects the
model parameters.  The harmonic weighting system described above makes the
optimal choice to mitigate the phase variability due to a particularly noisy
harmonic, but the TOA residuals are still of substantially greater magnitude
than would be expected from Poisson noise alone.  This leaves us with the task
of estimating the fit uncertainties in the presence of such noise.  These
uncertainties are crucial to our construction of timing models, as they are
needed to estimate the significance of fit components, such as frequency
derivatives and instantaneous frequency changes.

We noted in the previous section that we treat the TOA residuals as noise,
despite some of their variabilty arising from pulse shape changes that recur
in every outburst.  This is not a bad approximation: because we fit our models
separately for each outburst, correlations in the pulse profile variability
between outbursts are not relevant.  Furthermore, the power spectra of the TOA
residuals (see Fig.~\ref{fig:powerspecs} in \S\ref{sect:noiseproperties})
resemble the power-law noise spectra typically observed in actual red noise
processes, so treating it as such is reasonable.

We initially used the simplest possible timing model when fitting the TOAs of
each outburst in \tempo: a circular orbit and a constant frequency.  (Note
that we fit independent models for each outburst.  The uncertainties were too
large to phase connect between outbursts.)  When this simple model proved
insufficient to account for the phase residuals, we introduced a nonzero
$\dot\nu$ and instantaneous frequency changes, as needed.  However, there is a
danger of overfitting the data.  It is important to recognize that some of the
features in the residuals are probably pulse profile variability rather than
spin evolution.  We took great care in our attempts to distinguish between the
$\dot\nu$ measurements and the artifacts of intrinsic timing noise.

The colored nature of the timing noise in both harmonics is the primary
difficulty in the interpretation of the parameter fits.  \Tempo\ assumes that
the TOA uncertainties one gives it are white and approximately Gaussian, as is
the case of pure Poisson noise.  As a result, it systematically underestimates
the uncertainties in the fitted parameters in the presence of timing noise.
Red timing noise is particularly problematic, because it dominates on the long
timescales on which $\nu$ and $\dot\nu$ measurements depend.

Instead of adopting this white noise assumption of \tempo, we estimated
confidence intervals for $\nu$ and $\dot\nu$ using Monte Carlo simulations of
the timing residuals of each outburst.  After using \tempo\ to obtain the best
fit for a timing model, we calculated the power spectrum $P(f)$ of the timing
residuals that \tempo\ output.  This spectrum is a convolution of the true
noise spectrum and the sampling function; most notably, there is excess power
around 1~d, an artifact of \RXTE\ observations often being scheduled
approximately a day apart, and at the \RXTE\ orbital period of 96~min due to
Earth occultations.  We applied a low-pass filter to remove these peaks in an
attempt to approximate the underlying noise spectrum, $P'(f)$:
\begin{equation}
  P'(f) = P(f) \times \left[(1 - A) \exp(-f^2 \tau_{\rm c}^2) + A \right] .
\end{equation}
$\tau_{\rm c}$ gives the time scale for the low-pass cutoff.  $A$ gives the
fraction of high-frequency noise to let through, reproducing the
short-timescale scatter (principally but not entirely Poisson) that we
observed within each observation.  Typical values were 3~d and 10--20\%.

We then created thousands of sets of artificial phase residuals with the noise
properties of the filtered spectrum, $P'(f)$.  To reproduce the sampling
irregularities, we removed all points at times absent in the original data.
The parameters of the low-pass filter were tuned such that the mean power
spectrum of the resulting Monte Carlo residuals was as close as possible to
the original power spectrum, $P(f)$.  For each set of residuals, we measured
the frequency of the best linear fit (or, if our \tempo\ model fit for
$\dot\nu$, the frequency derivative of the best quadratic fit).  The standard
deviations of these measurements provided uncertainty estimates for the
respective parameter, this time more accurately accounting for the noise
spectrum.

\begin{sidewaysfigure*}
  \begin{center}
    \includegraphics[width=1\textwidth]{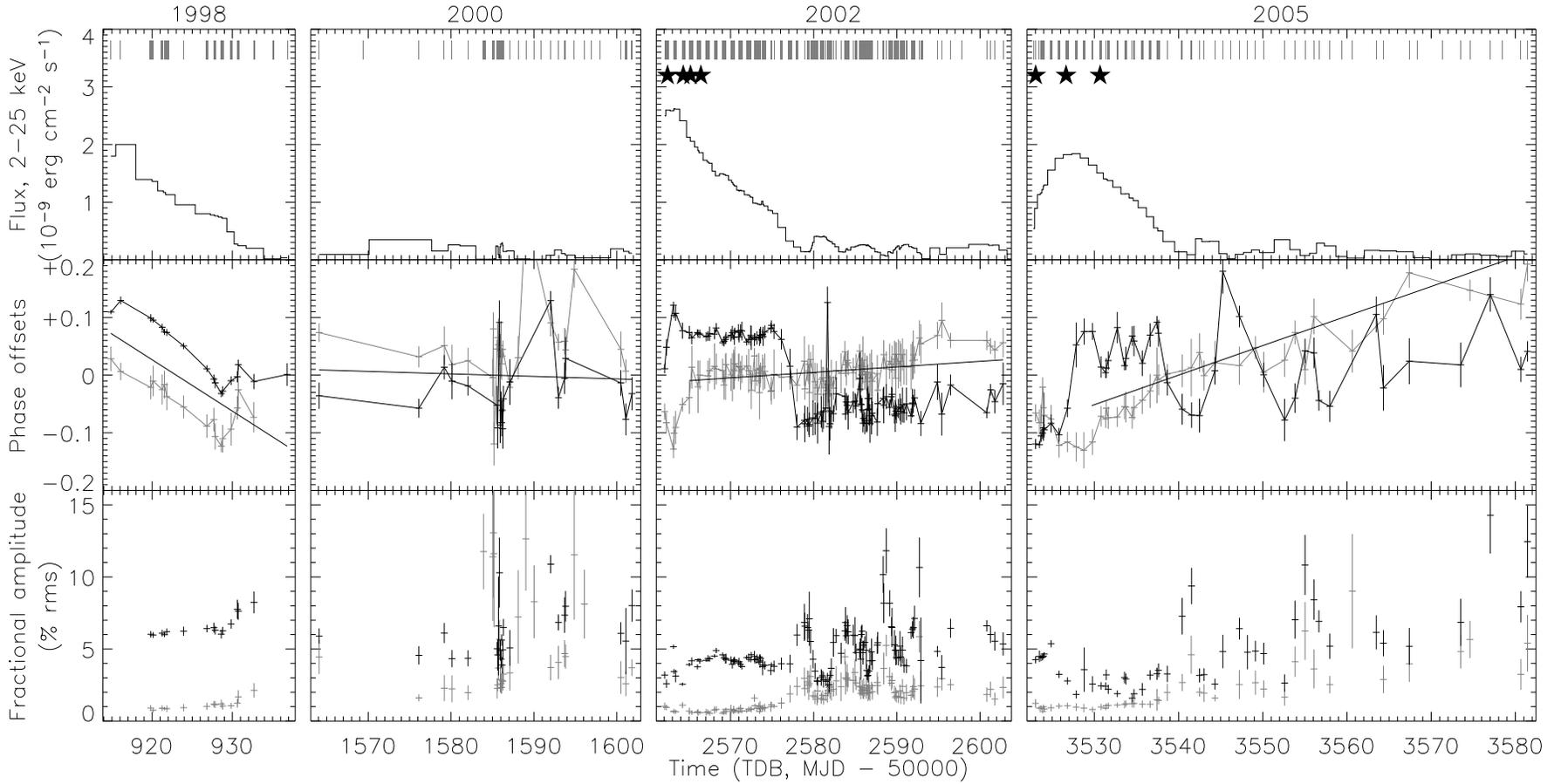}
  \end{center}
  \caption{ The light curves, phase residuals, and fractional amplitudes for
    all four outbursts.  The top panels show the background-subtracted light
    curves for each outburst.  The strips along the top of the graphs indicate
    the times of observations; stars indicate the times of thermonuclear X-ray
    bursts.  The second panels show show the phase residuals relative to a
    constant frequency of 400.97521025~Hz, with black points giving the phases
    of the fundamental and grey points indicating the second harmonic.
    Positive phases indicate pulse arrivals later than predicted by the phase
    model.  The error bars reflect the statistical errors only, as calculated
    in equation~(\ref{eq:sigmak}).  These data have been binned so that there
    is only one point per observation.  The black lines indicate the best-fit
    constant-frequency model for each outburst.  The bottom plot shows the
    fractional amplitudes of the fundamental and harmonic.
  \label{fig:bigplot}}
\end{sidewaysfigure*}

We relied solely on \tempo\ to calculate the uncertainties in the binary orbit
parameters.  While the intrinsic pulse profile noise spectrum is colored on
the timescale of days to weeks, the phase residuals are approximately white on
timescales equal to and shorter than the 2~hr orbital period.  The amplitude
of the short-timescale variability is roughly 1.5 times what one would expect
from counting noise alone, so we scaled our Poisson-derived phase
uncertainties accordingly when estimating the uncertainties of the orbital
parameter fits.  This rescaling makes \tempo's uncertainty estimates for the
orbital parameters reasonably accurate.  One important consistency check of
this simple approach worked nicely: the 1998, 2002, and 2005 measurements of
$P_{\rm orb}$ and $a_{\rm x} \sin i$, two parameters that should be the same
for each outburst at our level of accuracy, were indeed found to be constant,
with reduced $\chi^2$ statistics close to unity.

In deriving new binary and spin parameters, the new values sometimes differed
considerably from the parameters with which we initially folded the data for
TOA calculations.  If our orbital model improved substantially, we iterated
the above procedure, calculating new times of arrival for each 512~s interval
and refitting.  Because the orbit introduces a periodic frequency modulation
with amplitude $\Delta\nu = \nu_0 \cdot 2\pi a_{\rm x}\sin i / c P_{\rm orb} >
1 / 512$~s, an inaccurate orbital ephemeris can significantly reduce detection
strength.  In contrast, the spin frequency is remarkably stable through all
the observations, so there was no need to recalculate TOAs upon the relatively
minor revisions to the spin model.

\section{Results}
\label{sect:results}

The results of our pulse timing solutions are shown in
Figure~\ref{fig:bigplot}, which compares the light curves, phase residuals,
and fractional amplitudes for each outburst.  Inspecting the best-fit
frequency lines in the phase residual plots, it is clear that a constant pulse
profile attached to a constant-frequency rotator does not adequately describe
the observed residuals.  We consider five sources of phase residuals relative
to a best-fit constant-frequency model: Poisson timing noise, intrinsic pulse
profile noise, sudden and well-defined pulse profile changes, additional spin
frequency derivatives, and instantaneous frequency changes in the underlying
rotation of the star.  In this section, we will consider all these possible
contributions to the residuals and their relationships with each other and the
other properties of each outburst.

\subsection{Light curves of the outbursts}

\begin{figure}[t]
  \begin{center}
    \includegraphics[width=0.47\textwidth]{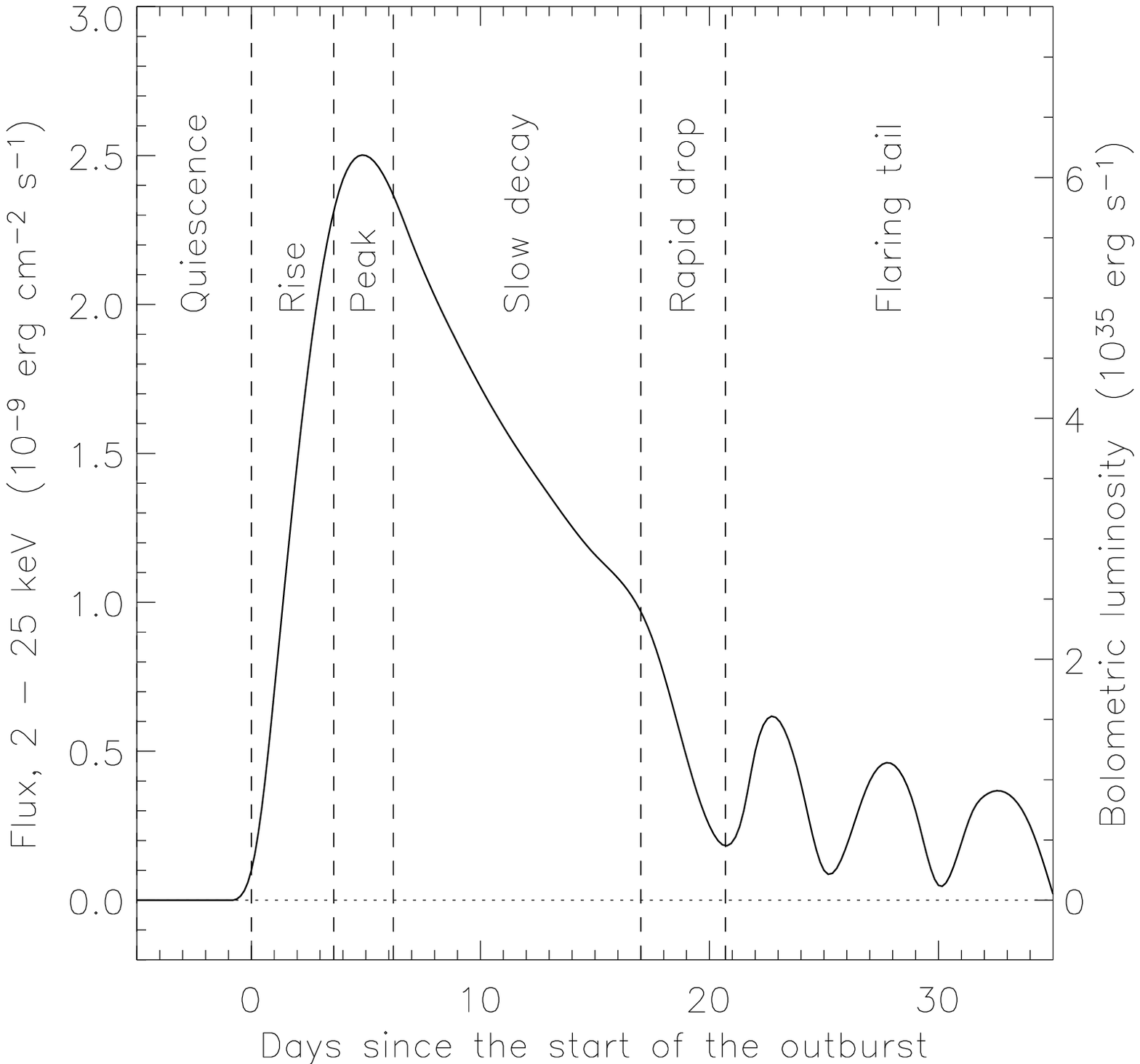}
  \end{center}
  \caption{ The anatomy of a typical outburst from \saxj.  The features of the
    light curve and their fluxes and timescales are similar to those observed
    during the 1998, 2002, and 2005 outbursts.  Bolometric luminosities assume
    a distance of 3.5~kpc and a correction of $L_{\rm bol} / L_{\rm 2-25\ keV}
    = 2.12$ \citep{Galloway06}.
  \label{fig:outburststages}}
\end{figure}

The light curves of each outburst are quite similar in shape.  We divide them
into four stages: the rise, which was only definitively captured in 2005 and
took $\approx$5~d; the short-lived peak at a 2--25~keV flux of
$(1.9$--$2.6)\times10^{-9}$\fluxunits, equal to a luminosity of
$(4.7$--$6.4)\times10^{35}$~erg~s$^{-1}$ using the distance of 3.5~kpc and
bolometric correction of $L_{\rm bol} / L_{\rm 2-25\ keV} = 2.12$ derived by
\citet{Galloway06}; a slow decay\footnote{Some authors
\citep[e.g.,][]{Cui98,Burderi06} refer to this part of the outburst as the
``exponential decay'' stage, based on the approximately exponential dimming of
the 1998 and 2002 outbursts.  However, the fall off of the 2005 outburst
during this stage is closer to linear, so we simply refer to it as the ``slow
decay'' stage to contrast it with the more rapid luminosity drop at its
conclusion.} in luminosity, lasting 10--15~d, until the source reaches
approximately $8 \times 10^{-10}$\fluxunits ($= 2.0\times
10^{35}$~erg~s$^{-1}$); and a sudden drop followed by low-luminosity flaring
as the outburst flickers out, with the timescale between flares on the order
of 5~d.  Figure~\ref{fig:outburststages} shows a cartoon of a typical outburst
from \saxj, with each of these stages labeled.

The \RXTE\ first collected high-resolution timing data from \saxj\ during the
1998 April outburst.  Data from the \RXTE\ All Sky Monitor (ASM) show that the
peak luminosity occurred approximately three days before the first PCA
observation.  (See Fig.~2 of \citealt{Galloway06} for a comparison of the ASM
and PCA light curves; note that its 1998 plot does not include two raster
scans analyzed here.)  Unfortunately, PCA observations stopped shortly after
the body of the outburst and do not sample the tail.

\saxj\ was discovered to be again in outburst when it emerged from behind the
Sun in 2000 January.  Coverage of the outburst was limited and included only
the outburst tail.  \citet{Wijnands01} comment on the erratic nature of the
flaring during the tail.  ASM data indicate that the peak occurred 2~weeks
prior to the first PCA observation and that we are observing the later, dimmer
stage of the flaring tail.  Comparison of the PCA data with the 2002 and 2005
outbursts suggests likewise.

The 2002 outburst, detected in mid-October and observed for the next two
months, was the brightest, had the best PCA coverage, and included the
detection of four extremely bright thermonuclear X-ray bursts during its peak.
Its light curve was very similar in shape to the 1998 outburst.

In 2005 June, \saxj\ was again in outburst.  This time, the detection preceded
the peak by a few days, providing a full sampling of the light curve.  This
outburst was somewhat dimmer, with a peak luminosity of only 70\% of the 2002
peak and a correspondingly shorter slow-decay stage.  The subsequent rapid
decay and flaring tail look quite similar to the other outbursts.

\begin{figure*}[t!]
  \begin{center}
    \includegraphics[width=1.0\textwidth]{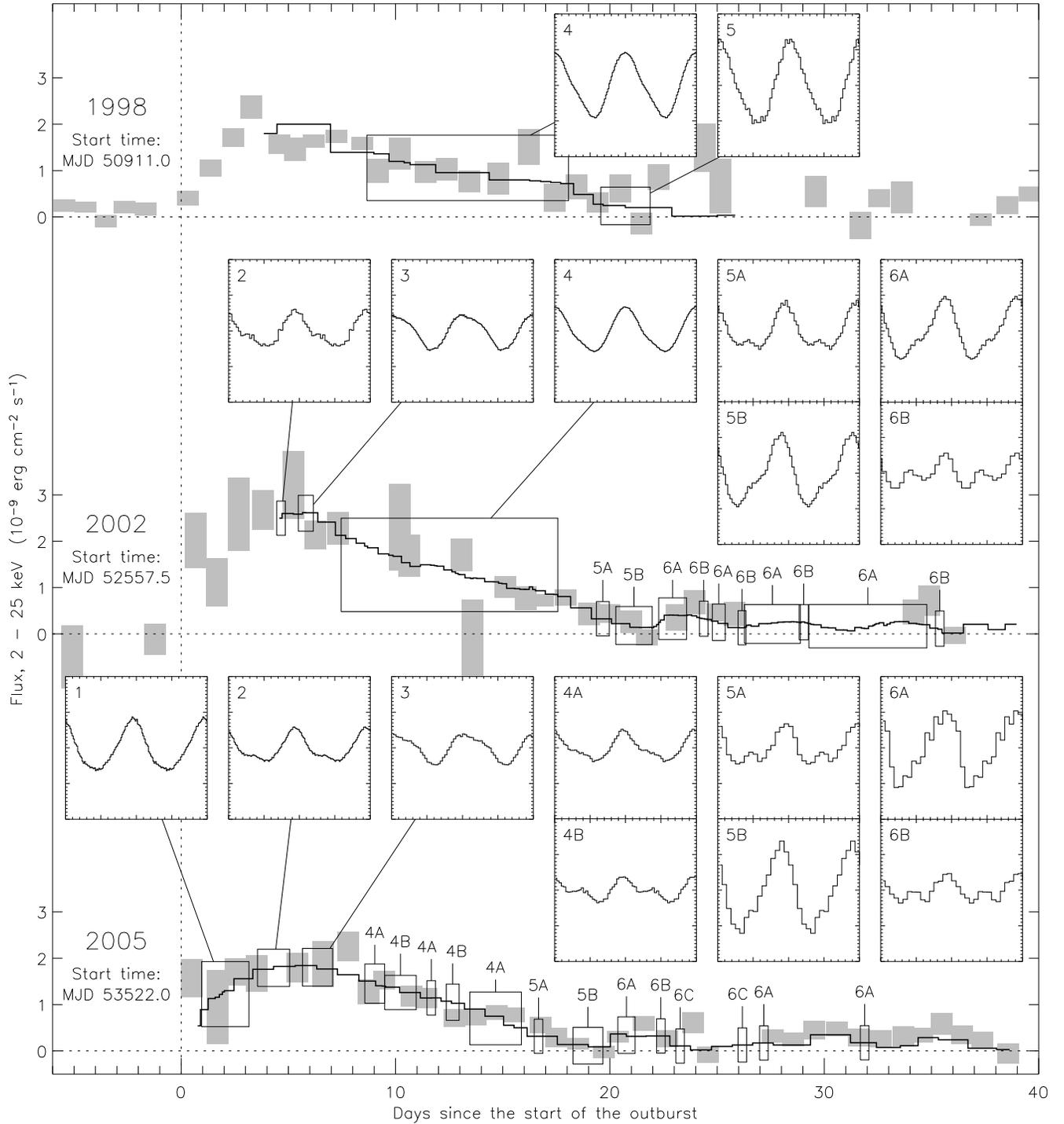}
  \end{center}
  \caption{ A comprehensive view of the 2--15~keV pulse profiles observed from
    \saxj.  Each pulse profile was calculated by folding the observations
    within the indicated time intervals using the best-fit constant-frequency
    model of each outburst, so any movement of the peaks reflects the phase
    offsets from the constant frequency.  The profiles are
    background-subtracted, normalized such that the phase bins have a mean
    value of unity, and plotted on 0.80--1.20.  Thus the plotted profiles
    accurately show the change in fractional amplitude during the outburst.
    The profiles are numbered according their position within the outburst: 1
    indicates the burst rise; 2, the beginning of the outburst maximum; 3, the
    end of the maximum; 4, the slow decay stage; 5, the steep luminosity drop
    marking the end of the main outburst; and 6, the flaring tail.  During
    some parts of the burst, two pulse profiles are present, with the source
    switching between them.  In these cases, we show both profiles and label
    the regions of the light curve in which they occurred accordingly.  The
    solid black line shows the fluxes from the PCA observations; the grey
    boxes show the fluxes from the ASM daily averages.
    \\
  \label{fig:pulseprofiles}}
\end{figure*}

\subsection{Characteristic pulse profile changes}
\label{sect:pulseprofiles}

Just as the light curves of each outburst were quite similar, the evolution of
the pulse profile during each outburst was remarkably consistent.
Figure~\ref{fig:pulseprofiles} illustrates the full range of pulse profiles
that we observed from \saxj.  In many instances, the similarity of the pulse
profiles between outbursts is quite striking.  In this section, we describe
how these profiles change throughout the outbursts.

We observed the outburst rise, labeled as profile~1 in
Figure~\ref{fig:pulseprofiles}, exclusively during the 2005 outburst.  The
profiles are smooth and asymmetric, with a slow rise followed by a more rapid
drop-off after the peak.  There is no sign of a second peak.

We observed the outburst maxima during 2002 and 2005.  The similarity of the
pulse profile evolution between the two outbursts is remarkable.  During the
first half of the maxima (labeled as profile set 2), the profiles show a
secondary bump lagging the main pulse.  Compared to the burst rise, the
fractional amplitude has decreased somewhat.  During the second half of the
maximum, the pulse becomes broader, subsuming the lagging secondary bump.
(See profile set~3.)  This change appears to be gradual: in both outbursts, a
mid-peak observation exhibited an intermediate pulse profile.

Profile set~4 shows the pulse profiles during the slow decay stages of the
outbursts.  The 1998 and 2002 profiles are quite similar: the pulses are
somewhat asymmetric, rising more steeply than they fall.  In both outbursts,
this pulse profile is very stable during the approximately 10~d of the decay
in luminosity.  During the 2005 outburst, this asymmetry is more pronounced,
and the profile varies between observations.  Initially, the pulse exhibited a
small lagging bump (profile~4A), quite similar to the pulse profile during the
first half of the outburst maximum.  The relative size of that bump varied
substantially, in some observations appearing as a small secondary peak
(profile~4B)\@.  Over the course of the decline, the source switched back and
forth between a double-peaked and single-peaked profile as indicated in the
figure.  A given state would typically be seen for two or three observations
(1--2~d) before switching to the other.

Profile set~5 covers the rapid drop in flux at the end of the outbursts.
During 1998, the pulse profile was quite stable and did not appreciably change
during this drop, although its fractional amplitude increased somewhat.  In
contrast, the 2002 and 2005 outbursts show a major pulse profile shift
concurrent with the drop in luminosity.  Prior to the drop, the pulses in
set~4 show a quick rise and a slower fall.  After the drop, the asymmetry of
the 2002 and 2005 profiles reverses: profiles~5B show a slow rise and a quick
drop.  In terms of harmonic components, these changes represent a shift in the
phase of the fundamental by approximately 0.15~cycles as it went from leading
the second harmonic to lagging behind it.  The phase of the harmonic did not
change.  In both outbursts, observations during the $\approx$2~d of rapid
luminosity decline reveal an intermediate stage in which the main pulse is
momentarily symmetric (profiles~5A)\@.  During this transition, small but
significant secondary pulses are present.

During the flaring tail of the outburst (profile set~6), the pulse profile
again showed substantial variability.  In 2002, the profile repeatedly
switched between an asymmetric pulse (profile~6A, identical to the pulse
profile at the end of the rapid dimming stage) and a double-peaked profile
(profile~6B)\@.  The double peaked pulse profile occurs principally (but not
exclusively) at the end of the flares, as their luminosity declines.  These
pulse profile changes are almost entirely the result of changing fractional
amplitudes of the harmonic components; the phase offset between the
fundamental and second harmonic remains for the most part constant.  A notable
exception occurs during the decay of the first flare at around MJD 52582.  At
this time the phase of the fundamental jumped by $\approx$0.2 cycles,
indicating a sudden lag of this amount behind its previous arrival time.  By
the next observation, less than two hours later, the phase residual of the
fundamental returned to its previous value.

The tail of the 2005 outburst is more chaotic.  The fractional amplitudes and
phases both exhibit strong red noise, producing a pulse profile that is
sometimes asymmetric with a slow rise and quick fall (6A); at other times
asymmetric with a quick rise and slow fall (6C; not shown, but basically just
the reverse of profile~6A); and in one instance clearly double-peaked (6B)\@.
The observations were sparse and generally short, so it was impossible to
better characterize the evolution of these pulse profile fluctuations.  The
flaring tail of the 2000 outburst was quite similar, with a highly variable
pulse profile that included double-peaked profiles and asymmetric single
pulses of both orientations.  We did not include it because the observations
were few and sparse.

\subsection{Noise properties of the timing residuals}
\label{sect:noiseproperties}

To measure the spin phase of \saxj\ using the formalism developed in
\S\ref{sect:toacalc}, we must characterize the variability of the harmonic
components.  This variability encompasses both the pulse profile changes
discussed in the previous section as well as any noise in the spin phase of
the star.

In our analysis of the phase residuals, we took into account the rms amplitude
of the intrinsic pulse profile noise in each harmonic, $\sigma^2_{k,{\rm
int}}$, defined in equation~(\ref{eq:noisesigmas}).  A casual glance at the
phase residuals of Figure~\ref{fig:bigplot} reveals that the magnitudes of
these fluctuations vary substantially between outbursts.
Table~\ref{tbl:noise} summarizes these amplitudes for each outburst and
compares them to the mean amplitudes of their Poisson noise,
$\left<\sigma_k^2\right>$$^{1/2}$.  These values are then used in
equation~(\ref{eq:weighting}).  For instance, in the 2002 outburst the
fundamental is more heavily weighted in measuring the spin phase than the
second harmonic, while in 2005 the opposite is true.

\begin{deluxetable*}{lccccccc}
\tabletypesize{\footnotesize}
\tablecolumns{8}
\tablewidth{0pt}
\tablecaption{Noise properties of the outbursts\label{tbl:noise}}
\tablehead{
  \colhead{} &
  \multicolumn{3}{c}{Fundamental} &
  \colhead{} &
  \multicolumn{3}{c}{Second harmonic}\\
  \cline{2-4} \cline{6-8}
  \colhead{\phantom{$\left<{}^2\right>^{1/2}$}} &  % for vertical spacing
  \colhead{$\left<\sigma_1^2\right>$$^{1/2}$} &
  \colhead{$\sigma_{1,{\rm int}}$} &
  \colhead{$\gamma_{\rm PLN}$} &
  \colhead{} &
  \colhead{$\left<\sigma_2^2\right>$$^{1/2}$} &
  \colhead{$\sigma_{2,{\rm int}}$} &
  \colhead{$\gamma_{\rm PLN}$}
}
\startdata
1998 Apr & 0.007 & 0.014 & $0.96\pm0.09$ & & 0.021 & 0.017 & $0.43\pm0.12$\\
2000 Feb & 0.023 & 0.052 &      ---      & & 0.022 & 0.039 &      ---     \\
2002 Oct & 0.012 & 0.016 & $0.67\pm0.09$ & & 0.023 & 0.027 & $0.51\pm0.09$\\
2005 Jun & 0.013 & 0.061 & $0.85\pm0.07$ & & 0.019 & 0.024 & $0.77\pm0.05$\smallskip
\enddata

\tablecomments{All phases are in cycles (i.e., fractions of the 2.5~ms spin
  period).  $\left<\sigma_k^2\right>$$^{1/2}$ gives the mean contribution of
  Poisson noise; $\sigma_{k,{\rm int}}$ is the amplitude of pulse profile
  variability in excess of the Poisson noise; and $\gamma_{\rm PLN}$ is the
  slope of the power law best fit to the spectrum of $\sigma_{k,{\rm int}}^2$.
  We did not attempt to estimate power law noise slopes for the 2000 outburst
  because of its low-quality data.}
\end{deluxetable*}

\begin{figure}[t!]
  \begin{center}
    \includegraphics[width=0.40\textwidth]{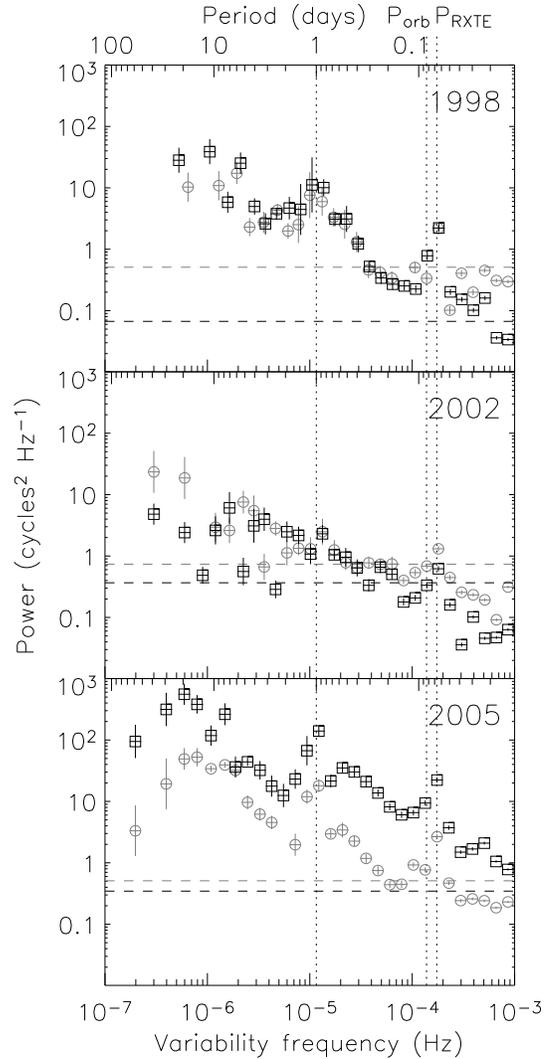}
  \end{center}
  \caption{ Power spectra of the phase residuals for the fundamental (black
    squares) and the second harmonic (grey circles) relative to the best-fit
    constant-frequency models.  (The 2002 model also includes a phase shift in
    the fundamental to account for the profile change at MJD 52577.)  The
    dashed lines show the power level due to counting statistics, a
    white-noise contribution proportional to $\left<\sigma_k^2\right>$.  The
    data points show the powers $P_k(f)$, from which we have subtracted the
    contribution of counting statistics.  These powers are normalized such
    that $\int_{10^{-7}\ {\rm Hz}}^{10^{-3}\ {\rm Hz}} P_k(f)\,df =
    \big<\sigma^2_{k,{\rm int}}\big>$, as defined in equation
    (\ref{eq:noisesigmas}).  The vertical dotted lines show the relevant time
    scales for the spectra: the 96~min and $\approx$1~d periodicities of the
    \RXTE\ observations, and the 121~min \saxj\ orbital period.
  \label{fig:powerspecs}}
\end{figure}

The scatter of the phase residuals between observations is generally greater
than the scatter within an observation, suggesting that the pulse profile
noise is red.  Power spectra of the phase residuals, shown in
Figure~\ref{fig:powerspecs}, confirm this.  We estimated these power spectra
using Fourier transforms of the residuals from equally spaced 512~s bins.
Here we have not attempted to deconvolve the uneven sampling periodicities at
1~d and 96~min due to the \RXTE\ observation schedule and orbit.  There are no
peaks at the 2~hr binary orbital period, indicating that the pulse profile is
independent of orbital phase.

The resulting noise powers are around 2~decades higher at long periods
($\approx$3~d or longer) than at short periods for 1998, and even more for
2005.  The 2002 outburst spectra exhibit less profile noise at long
timescales, but still are somewhat red.  Poisson statistics produce an
uncolored lower limit on noise.  This white noise dominates at timescales
shorter than the orbital period, except in the case of the particularly noisy
fundamental of 2005.  The spectra of the intrinsic profile noise (i.e., the
spectra after subtracting off the Poisson contribution) roughly followed a
power law noise spectrum, which we parametrized as $P_k(f) \propto
f^{-\gamma_{\rm PLN}}$.  The best-fit values of $\gamma_{\rm PLN}$, listed in
Table~\ref{tbl:noise}, varied from roughly 0.4 to 1.

\begin{figure}[t]
  \begin{center}
    \includegraphics[width=0.47\textwidth]{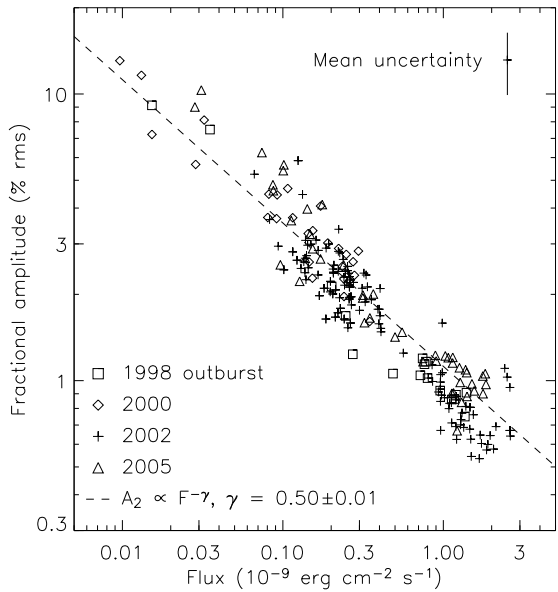}
  \end{center}
  \caption{ The fractional amplitude of the second harmonic scales with flux
    according to a power law of slope $-0.50 \pm 0.01$, shown by the dashed
    line.  Each point gives the mean amplitude and flux for a single
    observation.  The scatter is commensurate with the mean uncertainty in
    fractional amplitude, which is shown by the error cross in the upper
    right.
  \label{fig:ampvsflux}}
\end{figure}

\subsection{Fractional amplitudes of the harmonics}

In our time-domain discussion of the pulse profiles, an apparent trend is the
tendency of the pulses to become narrower, more asymmetric, or doubly peaked
--- generally speaking, to become less sinusoidal --- as the outburst's flux
decreases.  In the frequency domain, the relation is striking: the fractional
amplitude of the 802~Hz second harmonic, $r_2$, strongly anticorrelates with
the background-subtracted 2--25~keV flux, $f_{\rm x}$, as shown in
Figure~\ref{fig:ampvsflux}.  This power-law dependency has a slope of
$-0.50\pm0.01$.  The agreement with the data is excellent for such a simple
model, giving a reduced $\chi^2$ statistic of $\chi^2_\nu = 1.15$ with 1816
degrees of freedom.  It spans two and a half decades in luminosity and
includes every detected harmonic amplitude from all four outbursts.

In terms of the pulse profile, the second harmonic contributes in two ways.
If its peak is 45\degr\ out of phase with the peak of the fundamental, it will
produce an asymmetric pulse profile (e.g., profile~6A in
Fig.~\ref{fig:pulseprofiles}).  If it is in phase, a narrower primary pulse
with a small second peak will result (as in profile~6B).  If the components
are 90\degr\ out of phase, the profile will be profile~6B flipped, but we
never observed such a configuration.

To further understand the influence of flux on the pulse profile, we
decomposed the second harmonic's fractional amplitude into its asymmetric and
double-peaked components,
% Use subequations (in amsmath) with emulateapj; use mathletters with aastex.
\begin{subequations}  %\begin{mathletters}
\begin{eqnarray}
  r_{2,{\rm asym}} &=& r_2 \left|\sin 4\pi\psi\right| \quad \textrm{and} \\
  r_{2,{\rm dp}}   &=& r_2 \left|\cos 4\pi\psi\right| \, ,
\end{eqnarray}
\end{subequations}  %\end{mathletters}
where $\psi$ is the phase offset between the peaks of the two harmonics: $\psi
= (\phi_2 + \Delta\phi_2) - (\phi_1 + \Delta\phi_1)$.  The resulting plots
have substantially more scatter than Figure~\ref{fig:pulseprofiles} due to the
uncertainty of $\psi$, which is considerable, particularly at low fluxes.
However, they both roughly conform to the $r_2 \propto f_{\rm x}^{-1/2}$ power
law.  We conclude that the decrease in flux increases the asymmetry of the
pulses and the presence of secondary pulses in approximately equal measure.

In contrast, the fractional amplitude of the fundamental behaves
unpredictably.  During the slow-decay stage of 1998, it is unvarying and
strong, at a constant 5.5\% rms.  During this stage of 2002, it is weaker
(4\%) and somewhat variable; during 2005, it is weaker still and erratically
changing by up to a full percent between observations.  Its behavior is more
consistent in the tail.  In all outbursts, the fractional amplitude of the
fundamental varies widely, usually (but not always) having its maxima around
the peaks of the flares and its minima during the fading portion of the
flares.

For the most part, a pulse profile model only including the fundamental and
second harmonic adequately describes the folded profiles.  However, folding
long stretches of data does sometimes result in the detection of a third
harmonic with fractional amplitudes ranging up to $\approx$0.25\% rms.  We do
not reliably detect any higher harmonics.

\subsection{Upper limits on the subharmonics}

\begin{deluxetable}{ccccc}
\tabletypesize{\footnotesize}
\tablecolumns{5}
\tablewidth{0pt}
\tablecaption{Upper limits on subharmonics and half-integral harmonics
  \label{tbl:subharms}}
\tablehead{
  \multicolumn{2}{c}{Harmonic} &
  \multicolumn{3}{c}{Upper limit\tablenotemark{a} (\% rms)} \\
  \colhead{Factor} &
  \colhead{Hz\tablenotemark{b}} &
  \colhead{A} &
  \colhead{B} &
  \colhead{C}
}
\startdata
1/4 & \pz100.2 & 0.017 & 0.019 & 0.52\\
1/2 & \pz200.5 & 0.022 & 0.024 & 0.45\\
3/2 & \pz601.5 & 0.018 & 0.021 & 0.43\\
5/2 &   1002.4 & 0.026 & 0.024 & 0.42\smallskip
\enddata
\tablenotetext{a}{These background-corrected upper limits are quoted at the
95\% confidence level.  These limits result from combining all the
observations (column~A), combining only bright observations (B), and not
combining any observations (C).  See the text for more details.}
\tablenotetext{b}{Frequencies listed here are approximate.  The upper limits
were obtained using exact multiples of the best-fit constant-$\nu$ models.}
\end{deluxetable}

With some assumptions, we can strongly constrain the presence of subharmonics
and half-integral harmonics.  The most straightforward approach is to fold all
the observations using multiples of the best-fit frequency models from each
outburst.  The amplitude of the resulting profile will give an upper limit.
The resulting 95\% confidence upper limits are listed in column~A of
Table~\ref{tbl:subharms}.  However, this approach is only statistically valid
if the uncorrected fractional amplitude (i.e., the fractional amplitude
relative to the source counts {\em and} the background) is constant.  Clearly
this assumption is false.  Aside from the varying proportion of source
photons, the background-subtracted fractional amplitudes of the fundamental
and the second harmonic fluctuate throughout the outburst, spanning nearly an
order of magnitude in the tails of the burst.  There is no reason to believe
that a subharmonic would not fluctuate similarly.

Column~B of Table~\ref{tbl:subharms} takes the more moderate approach of only
folding together observations during which the 2--25~keV flux exceeds
$5\times10^{-10}$\fluxunits, thereby only including the main body of the
outbursts.  Background photons are thus a much smaller contribution, and the
fractional amplitudes of the observed two harmonics were relatively stable
during these times.  Nevertheless, we still are folding enough photons to
obtain very stringent upper limits: in the case of the 200~Hz subharmonic, we
get a 95\% confidence upper limit of 0.024\% rms.  We feel that these numbers
are our most reliable, not making unreasonable assumptions about the
fractional amplitude fluctuations.

For completeness, we also include the most conservative upper limits, which
make no assumptions whatsoever about the fractional amplitudes of the
subharmonics.  For instance, it would be possible in principle for the
subharmonic to be present only during a single observation and zero-amplitude
everywhere else.  To constrain the resulting upper limits at least somewhat,
we again only used observations during which the source was brighter than
$5\times10^{-10}$\fluxunits\ and that had at least $10^6$ counts.  These
single-observation limits are tabulated in column~C.

The stringent upper limits of column~B provide the best evidence yet that the
the spin frequency of the star is indeed 401~Hz.  If the star was spinning at
200.5~Hz, with two antipodal hot spots each emitting pulses to produce the
observed frequency, a 200.5~Hz subharmonic would almost certainly be present.

\subsection{Spin frequency measurements and constraints}

We initially performed the simplest possible fits to the phase residuals of
each outburst: constant-frequency models.  We did not include the data at the
very beginning of the 2002 and 2005 outbursts, where pulse profile changes
during the rise and peak obscure any variations in the phase.  We also
excluded the residuals during 2002's mid-outburst pulse profile change, but
included the residuals of the fundamental on both sides of the shift by using
different profile templates before and after it.

\begin{figure}[t]
  \begin{center}
    \includegraphics[width=0.47\textwidth]{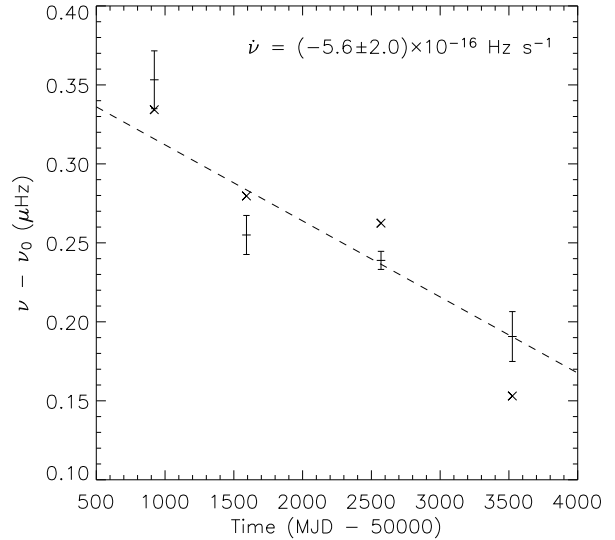}
  \end{center}
  \caption{ Constant-frequency measurements of the \saxj\ outbursts, showing
    the spin down of the star.  The frequencies are relative to $\nu_0 =
    400.97521000$~Hz.  The error bars are estimated using Monte Carlo
    simulations of phase residuals with the same noise properties as the
    actual outburst; they do not account for the uncertainty in the source
    position.  The $\times$'s mark what the frequencies would be if the fit
    source position differed from the actual position by $2\,\sigma$ along the
    ecliptic plane in the direction of increasing RA.  The same position error
    in the decreasing RA direction would move the frequency points by an equal
    amount in the opposite sense.
  \label{fig:spindown}}
\end{figure}

The resulting frequency measurements are shown in Figure~\ref{fig:spindown}
and summarized in Table~\ref{tbl:spinparams}.  These data clearly indicate
that the source is spinning down.  The probability that the actual spin
frequency is constant or increasing is less than $10^{-9}$ given the
uncertainty estimates.  These uncertainties do assume that our optical
position is exact, but the position error is excluded because its effects are
highly correlated; for instance, the 1998 and 2002 outbursts are six months
apart on the calender, so a position offset would produce equal and opposite
frequency displacements for the measurements from these outbursts.  There is
no position that would provide a statistically feasible constant or increasing
frequency.

\begin{deluxetable}{@{}lr@{ -- }lc@{}c@{}}
\tabletypesize{\footnotesize}
\tablecolumns{5}
\tablewidth{0pt}
\tablecaption{Best-fit constant frequencies, and their $\dot\nu$ upper
   limits\label{tbl:spinparams}}
\tablehead{
  \colhead{} &
  \multicolumn{2}{c}{Data included} &
  \colhead{\phantom{\tablenotemark{a}} $\nu - \nu_0$ \tablenotemark{a}} &
  \colhead{\phantom{\tablenotemark{b}} $\dot\nu$ \tablenotemark{b}}\\
  \colhead{} &
  \multicolumn{2}{c}{(MJD)} &
  \colhead{(\uHz)} & 
  \colhead{($10^{-14}$~Hz~s$^{-1}$)}
}
\startdata
1998 Apr & 50914.8 & 50936.9 & $0.371 \pm 0.018$ & $(-7.5, 7.3)$\\
2000 Feb & 51564.0 & 51601.9 & $0.254 \pm 0.012$ & $(-1.1, 4.2)$\\
2002 Oct & 52565.0 & 52602.8\tablenotemark{c} & $0.221 \pm 0.006$ & $(-1.3, 2.5)$\\
2005 Jun & 53529.6 & 53581.5 & $0.195 \pm 0.016$ & $(-0.5, 2.4)$\smallskip
\enddata
\tablenotetext{a}{The frequencies are relative to $\nu_{0} = 400.975210$~Hz.}
\tablenotetext{b}{95\% confidence intervals from the Monte Carlo simulations.}
\tablenotetext{c}{Excluding MJD 52575.7--52577.7.}
\end{deluxetable}

The linear fit through the measured frequencies is not particularly good: its
$\chi^2$ statistic is 9.7 with 2 degrees of freedom, yielding a probability of
about 1\% that the frequencies are drawn from a linear progression.  Once
again, changing the source position does not significantly change the result
or improve the fit, and changes in the position by more than the $1\,\sigma$
uncertainty along the ecliptic substantially worsen the linear fit.  To
estimate the uncertainties of the linear slope in light of this poor fit, we
rescaled the measurement errors such that reduced $\chi^2$ statistic would be
unity.  The resulting first-order spin derivative is $\dot\nu =
(-5.6\pm2.0)\times10^{-16}$~Hz~s$^{-1}$.  The large $1\,\sigma$ uncertainty
reflects the uncertainty in the slope of the frequency change, not in the
observation that the source is spinning down.  The probability that the
frequency is not decreasing is less than $10^{-9}$, as mentioned above, a
confidence of better than $6\,\sigma$.

Fitting second-order frequency models established that $\dot\nu$ is consistent
with zero during all the outbursts.  These measurements are particularly
sensitive to pulse profile variations, so care must be taken to not overfit
such features.  We again exclude the initial observations of the 1998, 2002,
and 2005 outbursts, because the pulse profile changes would induce large
non-zero $\dot\nu$ measurements that most likely do not reflect the spin of
the underlying neutron star.  Using \tempo\ to find the best-fit $\dot\nu$'s
and applying Monte Carlos to estimate their uncertainties, we arrived at the
95\% confidence intervals of Table~\ref{tbl:spinparams}.  Excluding the 1998
outburst, which had the shortest span of timing data and thus the most poorly
constrained $\dot\nu$, these 95\% confidence upper limits were all of order
$|\dot\nu| \lesssim 2.5\times10^{-14}$~Hz~s$^{-1}$.

The uncertainties in the measurement of the frequency preclude phase
connection between outbursts.  During the 920~d gap between the 2002 and 2005
outbursts, the 6~nHz frequency uncertainty from 2002 would accumulate to a
phase uncertainty of 0.5~cycles; the $2\times10^{-16}$~Hz~s$^{-1}$ uncertainty
in the long-term spin down would contribute 0.6~cycles.  Worse, these
estimates are best-case scenarios, since they assume that the spin down is
constant.

\begin{figure}[t]
  \begin{center}
    \includegraphics[width=0.49\textwidth]{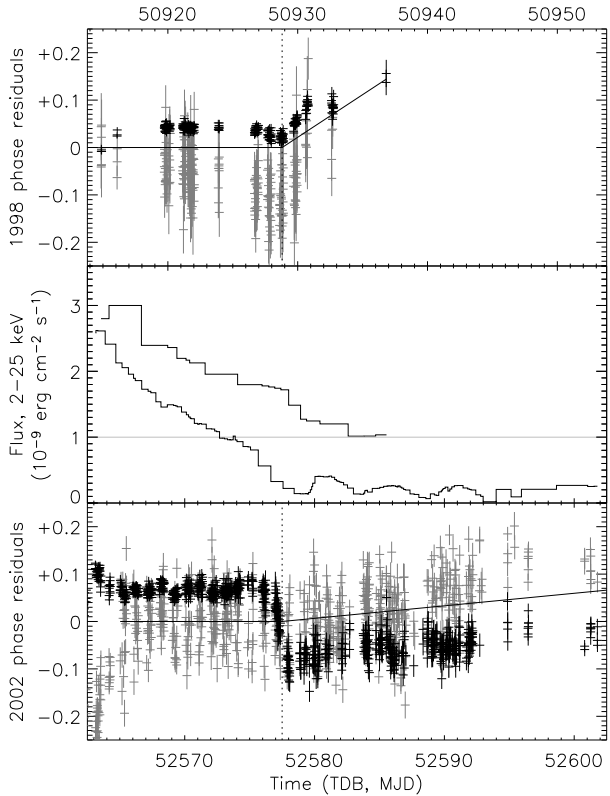}
  \end{center}
  \caption{ Comparison of the 1998 and 2002 glitch-like events.  The phase
    plots show the phase residuals relative to a constant-frequency model for
    the fundamental (black points) and the harmonic (grey points), binned such
    that there is one point per observation.  The black lines indicate the
    best timing models fit by \tempo.  The middle plot shows the 1998 and 2002
    light curves for comparison.  The 1998 light curve has been vertically
    offset by 1\fluxunits\ for clarity.  The data are displayed such that the
    apparent changes in frequency are aligned.  Notice that this alignment
    also has the effect of closely matching up the light curves.
  \label{fig:glitches}}
\end{figure}

During the 1998 and 2002 outbursts, we observed an abrupt change in the slope
of the phase residuals at the end of the main outburst.  We modeled these
apparent instantaneous changes of frequency by including frequency glitches in
our \tempo\ fits.  (While the \tempo\ glitch models are useful in describing
the data, we do not believe that we observed actual sudden changes in the spin
frequency of the star, a point discussed in detail in
\S\ref{sect:hotspotmotion}.)  Figure~\ref{fig:glitches} shows the phase
residuals of these outbursts and their best-fit glitch models.  These models
only employ an instantaneous change in frequency; including a phase jump or
introducing a $\dot\nu$ after the events did not significantly improve the
fits.\footnote{During the 2002 outburst, the absence of a phase jump refers
only to the second harmonic, which we believe is a better tracer of the
neutron star spin during this period of time (in agreement with the
conclusions of \citealt{Burderi06}).}

In both outbursts, these apparent frequency changes coincide with the sudden
drop in flux that marks the transition from the slow-decay stage to the
flaring tail stage.  At the same time, the fractional amplitudes of the
fundamental and harmonic increase, and, in the case of 2002, the pulse profile
change occurs.  (This pulse profile change, discussed earlier in
\S\ref{sect:pulseprofiles}, is apparent in Fig.~\ref{fig:glitches} as the
rapid advance of the fundamental phase.)  If we view the phase residuals with
respect to the pre-transition frequencies, as is the case in
Figure~\ref{fig:glitches}, the residuals following the transition skew upward,
indicating progressively increasing lags.  This effect is more pronounced in
1998, but its coverage is far better in 2002.  If we were to interpret these
changes in slope as abrupt spin frequency changes, they would represent drops
of 0.21~\uHz\ and 0.03~\uHz\ for 1998 and 2002, respectively.  (Again, we
consider this scenario unlikely; see \S\ref{sect:hotspotmotion}.)  If we
instead interpret them as the motion of a radiating spot, the drift rates
would be 6.5\degr~d$^{-1}$ and 1.0\degr~d$^{-1}$, retrograde.  The total
observed shifts between the start of the flaring tail and the loss of the
signal are substantial: 0.15~cycles (54\degr) in 1998 and 0.06~cycles
(22\degr) in 2002.  The data are not good enough to distinguish whether these
drifts are continuous.  For instance, it is possible that the hot spot made a
retrograde jump every time there was a flare.

We did not observe the main body of the 2000 outburst, so we cannot measure
whether the apparent frequency decreased when it entered the flaring tail
stage.  But if it did, and if the decrease in the apparent frequency was of
similar magnitude to that observed in 1998 and 2002, then including the main
body of the 2000 outburst would raise the overall frequency of the outburst
somewhat.  This correction might put it in line with the other frequency
measurements in Figure~\ref{fig:spindown}, reducing the large $\chi^2$
statistic of the constant-$\dot\nu$ fit.  Therefore we cannot conclude that
the change in the observed frequency from one outburst to the next is
incompatible with a linear progression.

During the 2005 outburst, the substantial pulse profile noise during the tail
prevented us from measuring a change in apparent frequency.  The uncertainty
in the measurement of the frequency during the tail was 0.03~\uHz, as
estimated using Monte Carlo simulations of the profile noise, and the phase
residuals jumped by as much as 0.1~cycles from one observation to the next.
If there was a smaller drift, as seen during the 2002 outburst, we would not
necessarily detect it.

\subsection{Evolution of the binary orbit}
\label{sect:porbderiv}

We fit the orbital parameters separately for each outburst.
Table~\ref{tbl:orbitparams} lists the results.  As expected, the values of
$a_{\rm x} \sin i$ and $P_{\rm orb}$ were consistent among the outbursts.  The
fit parameters $e\sin\omega$ and $e\cos\omega$ were consistent with zero.  We
used them to improve significantly on previous upper limits on the
eccentricity.

\begin{figure}[t]
  \begin{center}
    \includegraphics[width=0.47\textwidth]{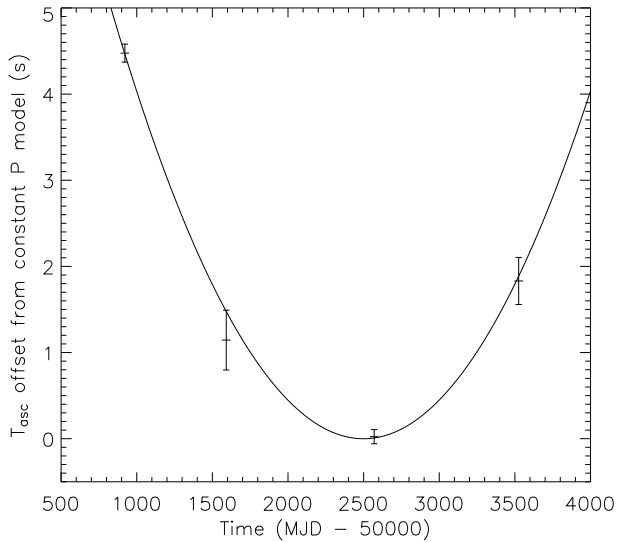}
  \end{center}
  \caption{ Measurement of an orbital period derivative.  The points show the
    observed times of ascending node, relative to the expected times for a
    constant period.  The $T_{\rm asc}$ of each outburst comes progressively
    later, indicating a period derivative of $(3.5\pm0.2) \times
    10^{-12}$~s~s$^{-1}$.
  \label{fig:tascadvance}}
\end{figure}

The measured time of ascending node advanced with each outburst, relative to
the times expected if the period was constant.  Figure~\ref{fig:tascadvance}
shows these $T_{\rm asc}$ residuals.  A quadratic provides a good fit ($\chi^2
= 1.01$ with a single degree of freedom), yielding a constant orbital period
derivative of $\dot{P}_{\rm orb} = (3.5\pm0.2)\times 10^{-12}$~s~s$^{-1}$ and
a significance of $15.6\,\sigma$.  In an independent analysis of the same
data, \citet{DiSalvo07b} report a consistent value for $\dot{P}_{\rm orb}$.
They derive a smaller uncertainty and larger $\chi^2$, most likely reflecting
an underestimate of the orbital phase measurement errors.

Table~\ref{tbl:paramsummary} summarizes all the parameters for the pulse
timing of \saxj.

\begin{deluxetable*}{lcclrr}
\tabletypesize{\footnotesize}
\tablecolumns{6}
\tablewidth{0pt}
\tablecaption{Binary parameter measurements from each
  outburst\label{tbl:orbitparams}}
\tablehead{
  \colhead{} &
  \colhead{$P_{\rm orb}$} &
  \colhead{$a_{\rm x} \sin i$} &
  \colhead{$T_{\rm asc}$} &
  \colhead{$e \sin\omega$} &
  \colhead{$e \cos\omega$}\\
  \colhead{Outburst} &
  \colhead{(s)} &
  \colhead{(light-ms)} &
  \colhead{(MJD, TDB)} &
  \colhead{($10^{-6}$)} & 
  \colhead{($10^{-6}$)}
}
% See the ``Improved measurement of the orbital period of SAX J1808'' section
% of 2005-11-01-TempoProject for the source of these numbers.
\startdata
1998 Apr & 7249.1553(18)   & 62.8080(46)\pz & 50921.7584194(12) & $ -60\pm64$ & $-86\pm64$\\
2000 Feb  &       ---       &       ---      & 51591.8019861(40) &  ---\pz\pz  &  ---\pz\pz\\
2002 Oct  & 7249.1565(6)\pz & 62.8147(31)\pz & 52570.0186514(9)  & $   8\pm57$ & $ 41\pm57$\\
2005 Jun  & 7249.1547(24)   & 62.8282(109)   & 53524.9944192(32) & $-173\pm83$ & $ 53\pm83$\smallskip
\enddata
\tablecomments{We excluded the 2000 outburst when calculating everything but
$T_{\rm asc}$ because its data were noisy and sparse.}
\end{deluxetable*}

\begin{deluxetable}{@{} lr @{}}
\tabletypesize{\footnotesize}
\tablecolumns{2}
\tablewidth{0pt}
\tablecaption{Combined timing parameters for \saxj
  \label{tbl:paramsummary}}
\startdata
\hline\hline\\[-1.5 ex]
Orbital period, $P_{\rm orb}$ (s) \tablenotemark{a} &
  7249.156961(14)\\
Orbital period derivative, $\dot{P}_{\rm orb}$ ($10^{-12}$~s~s$^{-1}$) &
  3.48(23)\\
Projected semimajor axis, $a_{\rm x} \sin i$ (light-ms) &
  62.8132(24)\\
Time of ascending node, $T_{\rm asc}$ (MJD, TDB) &
  52499.9602477(10)\\
Eccentricity, $e$ (95\% confidence upper limit) &
  $< 1.2\times 10^{-4}$\\
Spin frequency, $\nu$ (Hz) \tablenotemark{a} &
  400.975210240(11)\\
Spin frequency derivative, $\dot\nu$ ($10^{-16}$~Hz~s$^{-1}$) &
  $-$5.6(2.0)\smallskip
\enddata
\tablenotetext{a}{$P_{\rm orb}$ and $\nu$ are specified for the time $T_{\rm asc}$.}
\end{deluxetable}

\section{Discussion}
\label{sect:discussion}

Our analysis of multiple outbursts from \saxj\ allows us to greatly improve
our understanding of the behavior of this low-mass X-ray binary.  By comparing
the observed frequency from each outburst, we can see the long-term spin down,
which is too small to be detectable from a single outburst.  Comparison of
the pulse profiles from each outburst lead us to conclude that we are seeing
characteristic, repeated profile changes as the outbursts progress, rather
than a purely random noise process.  Finally, fitting of the orbital
parameters over the seven years of observation provides a greatly improved
orbital ephemeris.

\subsection{Long-term spin down}

By observing the mean spin frequency of each outburst, we found that \saxj\ is
spinning down at a rate of $\dot\nu = (-5.6\pm2.0)\times10^{-16}$~Hz~s$^{-1}$.
This spin down results in a loss of rotational energy at a rate of $\dot{E} =
4\pi^2 I \nu\dot\nu = 9\times10^{33}$~erg~s$^{-1}$, assuming a canonical value
of $I = 10^{45}$~g~cm$^2$ for the neutron star (NS) moment of inertia.

Most of this spin down occurs during X-ray quiescence; accretion torques
during the outbursts play a minimal role.  Over the seven years of our
observations, the mean outburst frequency decreases by $\nu_{2005} -
\nu_{1998} = -0.18\pm0.02\ \mu{\rm Hz}$.  Let us suppose that this frequency
change happens only during the X-ray outbursts (which have a duty cycle of
$\lesssim$5\%).  Since the outburst light curves are quite similar, it is
reasonable to presume that each would contribute roughly the same frequency
shift, thus splitting this frequency change into three equal steps.  If the
spin down is due to a constant $\dot\nu_{\rm outburst}$ that acts during the
$\approx$20~d of each outburst,\footnote{In reality, $\dot\nu$ would almost
certainly {\em not} be constant during as the accretion rate varies, but for
argument's sake we make the most conservative assumptions possible.  A varying
$\dot\nu$ would require that it be sometimes greater than the value from
equation~(\ref{eq:outburstonlyfdot}), making it even less plausible that it
would escape detection.} then
\begin{equation}
  \dot\nu_{\rm outburst}
     \approx \frac{-0.18\ \mu{\rm Hz}}{3 \times 20\ {\rm d}}
     = -3.5\times10^{-14}\textrm{ Hz s}^{-1} \, .
  \label{eq:outburstonlyfdot}
\end{equation}
By contrast, we were able to set stringent (95\% confidence) upper limits of
$|\dot\nu| \lesssim 2.5\times10^{-14}$~Hz~s$^{-1}$ during the outbursts
(Table~\ref{tbl:spinparams}).  We conclude that the spin down is dominated by
torques exerted during X-ray quiescence.

We will thus consider three possible sources of torque during quiescence:
magnetic dipole radiation, the expulsion of matter by the magnetic field
(i.e., the propeller effect), and gravitational radiation.  In general, we
assume that all three mechanisms contribute addititively to the observed spin
down of \saxj,
\begin{equation}
  N_{\rm obs} = N_{\rm dipole} + N_{\rm prop} + N_{\rm gr} .
\end{equation}
We discuss each below.

\subsubsection{Magnetic dipole torque}

A spinning dipolar magnetic field will produce a significant spin down during
quiescence for the $10^8$~G field strengths expected for a millisecond pulsar.
Relativistic force-free MHD models of pulsar magnetospheres by
\citet{Spitkovsky06} give a torque of $N_{\rm dipole} = -\mu^2 (2\pi\nu/c)^3
(1 + \sin^2\alpha)$, where $\mu$ is the magnetic dipole moment and $\alpha$ is
the angle between the magnetic and rotational poles.  Pulse profile modeling
of the 1998 outburst by \citet{Poutanen03} suggests that the magnetic hot spot
is not far from the rotational pole, separated by an angle of 5--20\degr.
While other effects might also contribute to the spin down, the rotating
magnetic field will always be present and provides an upper limit on the
dipole moment:
\begin{eqnarray}
  \mu & < & 0.77\times10^{26}
    \left(1 + \sin^2\alpha\right)^{-1/2}
    \nonumber\\ & & \times
    \left(\frac{I}{10^{45}\ {\rm g\ cm^2}}\right)^{1/2}
    \left(\frac{\nu}{401\ {\rm Hz}}\right)^{-3/2}
    \nonumber\\ & & \times
    \left(\frac{-\dot\nu}{5.6\times10^{-16}\ {\rm Hz\ s^{-1}}}\right)^{1/2}
    \textrm{ G cm}^3\, .
  \label{eq:spindowndipole}
\end{eqnarray}
For $\alpha = 15\degr$, this upper limit on the dipole is
$0.75\times10^{26}$~G~cm$^3$, yielding a field strength of roughly $B =
1.5\times10^8$~G at the magnetic poles.\footnote{The \citet{Spitkovsky06}
formula for $N_{\rm dipole}$ differs substantially from the classically
derived torque due to a rotating dipole in a vacuum, $N_{\rm vac} =
\frac{2}{3} \mu^2 (2\pi\nu/c)^3 \sin^2\alpha$, especially for small $\alpha$:
for $\alpha = 15\degr$, the derived limit is approximately one fifth of the
vacuum value.}  We emphasize that this upper limit on the magnetic field is
for a purely dipolar field.  The presence of higher-order multipoles would
require a stronger field at the NS surface to produce the observed $\dot\nu$.
This field estimate is consistent with the limits implied by accretion physics
(see \S\ref{sect:BFieldConstraints}).

If magnetic dipole torque is a significant contributor to the spin down of
\saxj, then the source may behave like a rotation-powered pulsar during
quiescence, producing radio pulsations and a particle wind.  The heating of
the companion by a particle wind has been invoked as an explanation of why the
companion is significantly brighter than expected in the optical.
\citet{Burderi03} predicted a dipole moment of $\mu = 5\times10^{26}$~G~cm$^3$
based on the optical observations, somewhat higher than our approximate upper
limit on $\mu$, but most likely within the uncertainties of the model.  A
similar analysis by \citet{Campana04} found the needed irradiation luminosity
to be $L_{\rm x} = (4^{+3}_{-1})\times10^{33}$~erg~s$^{-1}$, compatible with
the observed $\dot{E} = 9\times10^{33}$~erg~s$^{-1}$ loss of rotational
energy.  No radio emission has been detected during quiescence.  The upper
limits of 0.5~mJy \citep{Gaensler99, Burgay03} are not particularly
constraining.

The X-ray luminosities of isolated millisecond pulsars, for which magnetic
dipole radiation is the primary spin-down mechanism, shows a strong
correlation with their rates of rotational energy loss.  From the tables
compiled in \citet{Zavlin06} and \citet{Cameron07}, the 5--10~keV X-ray
luminosity goes as $L_{\rm x} \propto \dot{E}^{1.13}$ with less than a quarter
decade of scatter.  Based on this empirical relation, we would expect a
quiescent luminosity for \saxj\ of $5\times10^{30}$~erg~s$^{-1}$.  However,
this prediction is a factor of ten lower than the observed quiescent fluxes of
$8\times10^{31}$~erg~s$^{-1}$ and $5\times10^{31}$~erg~s$^{-1}$
\citep{Campana02, Heinke07}, suggesting other mechanisms for quiescent
emission are at work.

\subsubsection{Magnetic propeller torque}
\label{sect:propspindown}

The propeller effect offers another possible explanation for the observed spin
down during quiescence.  If the Keplerian corotation radius ($r_{\rm co} = [GM
/ 4\pi^2\nu^2]^{1/3}\approx 31$~km) is less than the magnetospheric radius
$r_0$, at which point the infalling matter couples to the magnetic field, then
the magnetic field will accelerate the matter, possibly ejecting it from the
system \citep{Illarionov75}.  The torque exerted on the neutron star by
propeller ejection of matter at a rate $\dot{M}_{\rm ej}$ depends on the
details of the interaction between the pulsar magnetosphere and the accretion
disk.  However, we can parametrize this torque as
\begin{eqnarray}
  N_{\rm prop} &=& -n \dot{M}_{\rm ej} (GMr_0)^{1/2} \nonumber\\
               &=& -n (r_0/r_{\rm co})^{1/2}
                   \dot{M}_{\rm ej} (GM r_{\rm co})^{1/2} \,,
\end{eqnarray}
where the detailed physics determines the dimensionless torque $n$, which is
zero for $r_0=r_{\rm co}$ and of order unity for $r_0\gtrsim 1.1\; r_{\rm co}$
\citep{Eksi05}.

We can then roughly estimate the rate at which matter would need to be ejected
from the system during quiescence to account for the observed spin down:
\begin{eqnarray}
  \dot{M}_{\rm ej} & < & -2.3\times10^{-12} \; n^{-1}
    (r_0 / r_{\rm co})^{-1/2}
    \nonumber\\ & & \times
    \left(\frac{I}{10^{45}\ {\rm g\ cm^2}}\right)
    \left(\frac{M}{1.4\ M_\sun}\right)^{-2/3}
    \left(\frac{\nu}{401\ {\rm Hz}}\right)^{1/3}
    \nonumber\\ & & \times
    \left(\frac{-\dot\nu}{5.6\times10^{-16}\ {\rm Hz\ s^{-1}}}\right)
    \ M_\sun\textrm{ yr}^{-1} \,.
\end{eqnarray}
As a consistency check, we note that this upper limit does not exceed the
predicted long-term mass transfer rate for the binary, $1\times10^{-11}\
M_\sun$~yr$^{-1}$, which is driven by gravitational radiation emission due to
the binary orbit \citep{Bildsten01}.  Indeed, not all the mass lost by the
donor star will necessarily reach the pulsar magnetosphere during quiescence
and be propelled outward; most of it would queue up in the accretion disk and
later reach the NS during an outburst.  \citet{Galloway06} found that the mass
transfer is roughly conservative, albeit with enough uncertainty that
propeller mass loss as large as the above $\dot{M}_{\rm ej}$ limit is not
ruled out.

Even if propeller spin down provides the dominant quiescent torque, the
resulting ejection of matter from the system would not greatly affect the
binary orbit.  The timescale for propeller spin down is proportional to the
timescale for the ejection of mass: $\dot{P}_{\rm orb}/P_{\rm orb} \propto
\dot{M}_{\rm ej}/M_{\rm c}$, where $M_{\rm c} \approx 0.05\ M_\sun$ is the
mass of the companion.  Applying the above $\dot{M}_{\rm ej}$ gives $M_{\rm c}
/ \dot{M}_{\rm ej} = 20$~Gyr, far longer than the observed orbital evolution
timescale of $P_{\rm orb} / \dot{P}_{\rm orb} = 66$~Myr.  More refined
calculations using the arguments of \citet{Tauris06} yield a propeller
timescale of 6~Gyr, still far too large.  Clearly there are other
contributions to the orbital evolution; we discuss some in
\S\ref{sect:discussporb}.

\subsubsection{Gravitational radiation torque}

A variety of mechanisms have been proposed in which rapidly rotating neutron
stars can develop mass quadrupoles that give rise to gravitational radiation
from the neutron star itself.  These mechanisms include $r$-mode instabilities
\citep{Wagoner84, Andersson99}, accretion-induced variations in the density of
the NS crust \citep{Bildsten98, Ushomirsky00}, distortion of the NS due to
toroidal magnetic fields \citep{Cutler02}, and magnetically confined mountains
at the magnetic poles \citep{Melatos05}.  The loss of angular momentum due to
gravitational radiation has been suggested as a mechanism to explain the
absence of observed pulsars with spin frequencies faster than $\approx$730~Hz
\citep{Chakrabarty03,Chakrabarty05} and makes millisecond pulsars a target for
interferometric gravitational wave detectors.

The mass quadrupole moment of the star, $Q$, determines the torque produced by
gravitational radiation: $N_{\rm gr} = -\frac{32}{5}GQ^2(2\pi\nu/c)^5$.  For
our measured $\dot\nu$, this sets an upper limit of
\begin{eqnarray}
  Q & < & 4.4\times10^{36}
    \left(\frac{I}{10^{45}\ {\rm g\ cm^2}}\right)^{1/2}
    \left(\frac{\nu}{401\ {\rm Hz}}\right)^{-5/2}
    \nonumber\\ & & \times
    \left(\frac{-\dot\nu}{5.6\times10^{-16}\ {\rm Hz\ s^{-1}}}\right)^{1/2}
    \textrm{ g cm}^2 \, ,
\end{eqnarray}
or $Q\lesssim 10^{-8}\,I$.  The strain amplitude of the resulting
gravitational waves, averaged over all NS orientations, is $h_c = 115 G \nu^2
Q / dc^4$ \citep{Brady98}, giving a characteristic strain at Earth of $h_c =
6\times10^{-28}$.  This strain is undetectable by current or planned
gravitational wave experiments.  For Advanced LIGO, with a strain sensitivity
of $\sim$$3\times10^{-24}$~Hz$^{-1/2}$ in the 100--400~Hz range
\citep{Fritschel03}, even a search using an accurate phase model would require
years of integration time.  Note that the dependence of $N_{\rm gr}$ on the
$\nu$ is very strong, so it is quite possible that gravitational wave emission
produces larger spin downs in faster ($\approx$700~Hz) rotators.

\subsection{Pulse profile variability}

The evolution of the pulse profile is clearly not purely stochastic.  With
multiple outbursts, we are able to note for the first time that the pulse
profile seems to take on similar shapes at similar times in the outbursts, as
illustrated in Figure~\ref{fig:pulseprofiles}.  These characteristic changes
in the pulse profiles suggest that the emitting regions of the NS are changing
shape and position as the outbursts progress.  The consistency of these
changes, along with the consistency of the outburst light curves, suggests
that as the accretion disk empties onto the star, the geometry of the disk,
the accretion funnels, and the resulting hot spots evolve in a similar manner
for each outburst.

The most striking example of ordered pulse-profile evolution is the strong
relationship between the harmonic content and luminosity: $r_2 \propto
L^{-1/2}$.  Given the complexity of the system, its abidance by such a simple
model is quite surprising.  \saxj\ is not alone in this behavior.  At least
two other millisecond pulsars, IGR~J00291$+$5934 and XTE~J1807$-$294, exhibit
similar inverse correlations between the amplitude of their second harmonics
and luminosity (Hartman~et~al. 2007, in prep.).

One possible explanation is recession of the accretion disk as the accretion
rate drops, revealing the star's previously occulted second hot spot.  For
rapidly rotating pulsars, partial occultation of the star by the accretion
disk will be common.  Assuming a mass of 1.4~$M_\sun$, the co-rotation radius
of \saxj\ is $r_{\rm co} = 31\textrm{ km}\approx 3R$, where $R$ is the NS
radius.  Following the standard pulsar accretion model
\citep[e.g.,][]{Ghosh79b}, the inner edge of the accretion disk will be at
roughly the Alfv\'en radius: $r_0 \approx r_A \equiv (2GM)^{-1/7}
\dot{M}^{-2/7} \mu^{4/7}$.  This truncation radius must be at $r_0 < r_{\rm
co}$ for infalling matter to reach the NS surface.  There are clear problems
with the application of this model, which was developed for higher-field
pulsars with $r_0 \gg R$: the width of the transition region in which the
magnetic field becomes dominant is on the same order as its distance to the
star, muddling the definition of a truncation radius.  Nevertheless, this
simple model is still qualitatively instructive.

Since $r_{\rm co} \approx 3R$, neutron stars in systems with inclinations $i
\gtrsim 70\degr$ will always be partially occulted during outburst.  During
the outbursts of \saxj, the peak fluxes at which the pulses are most
sinusoidal are roughly a factor of 10 greater than the low fluxes at which the
harmonics are more prevalent (cf. Fig.~\ref{fig:ampvsflux}).  As a result, the
Alfv\'en radius will increase by a factor of $r_{\rm A,tail} / r_{\rm A,peak}
\approx 10^{2/7} \approx 2$ as the source dims.  Because the maximum Alfv\'en
radius is $\approx$$3R$ during accretion, the radius during the peak of the
outbursts must be $r_{\rm A,peak} \lesssim \frac{3}{2}R$.  At this separation,
the star will be partially occulted if $i \gtrsim 45\degr$.  Thus the degree
of occultation will depend on $\dot{M}$ for a wide range of inclinations.  For
$45\degr \lesssim i \lesssim 70\degr$, the disk will partially occult the NS
above some critical $\dot{M}$.  For $i \gtrsim 70\degr$, the NS will always be
partially occulted, with the degree of occultation increasing as $\dot{M}$
increases.  Pulse profile modeling by \citet{Poutanen03} suggests that the
system is at an inclination of $i > 65\degr$.

The observations that show clearly double-peaked pulse profiles happen
exclusively in the final, flaring tail stage of the outbursts, typically
during the fading portion of a flare.  In view of this model, one could
imagine that the accretion disk is most recessed as the flares fade.  One
difficulty with this model is that the increased $r_2$ observed at low
luminosities is not solely due to the appearance of doubly peaked pulse
profiles; many profiles in this regime show single pulses, but with
substantially greater asymmetry than typically seen at higher luminosities.

Another possible cause is the expansion of the hot spots during high accretion
due to diffusive effects.  Simulations of accretion flows by
\citet{Romanova04} demonstrate that as the fluence increases, the
cross-sections of the accretion funnels grow.  Modeling by \citet{Muno02}
establishes that the harmonic content of the pulsations decreases as the size
of the hot spot increases.

\subsection{Motion of the hot spot}
\label{sect:hotspotmotion}

During the 1998, 2002, and 2005 outbursts, we observed clear trends in the
phase residuals that suggest that the emitting regions do not remain at a
fixed longitude.  During 2002 and 2005, an abrupt phase change in the
fundamental at the end of the main body of the outburst produces an advance of
the pulse peak that corresponds to a shift of the hot spot by
$\approx$50\degr\ eastward.\footnote{For a more natural description, we adopt
the Earth-based convention of longitude: earlier pulse arrivals $\equiv$
prograde hot spot motion $\equiv$ eastward shift, and vice versa.}  These
shifts are simultaneous with and occur on the same 3--4~d timescale as the
sudden drops in luminosity at the end of the main outbursts.  During 1998 and
2002, the phase residuals of both harmonics begin gradually increasing during
the flaring tails of the outburst, corresponding to a westward drift of the
hot spots.  Motion of the hot spot has also been suggested to explain phase
residuals in GX~1$+$4 and RX~J0812.4$-$3114 \citep{Galloway01} and
XTE~J1814$-$338 \citep{Papitto07}.

These trends in the phase residuals almost certainly represent motion of the
observed hot spot rather than frequency glitches.  Glitches are rapid changes
in the spin frequency of the NS due to imperfect coupling between the crust
and more rapidly rotating, superfluidic lower layers
\citep[e.g.,][]{Anderson75}.  This interaction occurs well below the accretion
layer, and it would not be expected to coincide with or have the same
timescale as rapid changes in the accretion rate.

When discussing the motion of the hot spots, the longitudes of the magnetic
poles provide natural meridians from which to measure phase.  Since their
movement would require the realignment of currents in the core and crust, the
magnetic poles remain at fixed positions for timescales far longer than the
outbursts.  The suppression of regions of the field due to accretion also
occurs on long timescales \citep{Cumming01}.

For high-field pulsars, the magnetospheric radius is far from the star, and
the accretion column follows field lines that reach the NS surface near the
magnetic pole.  This is not necessarily the case for low-field pulsars.  A
closer accretion disk will intersect more curved field lines, which terminate
farther from the poles.  In the previous section, we described how the
Alfv\'en radius can move outward from roughly $1.5R$ to $3R$ as the accretion
rate drops.  As the disk recesses, it will intersect decreasingly curved field
lines that are rooted closer to the poles, causing the hot spots at the bases
of the accretion columns to also approach the poles.

This simple picture can explain the observed phase shift as the luminosity
rapidly drops during the end of the 2002 and 2005 outbursts.  In both cases,
the luminosity decreases by about a factor of 4.  A change in $\dot{M}$ by
this magnitude would cause the Alfv\'en radius to move outward by a factor of
1.5 and the inner edge of the accretion disk to move outward by a similar
amount.  This change will almost certainly cause material removed from the
inner edge of the disk to attach to a different set of field lines, with the
larger radius favoring lines that attach closer to the pole.  If the hot spot
tends to be to the west of the pole, as seen in \citet{Romanova04} for a
magnetic pole an angle of $\alpha = 30\degr$ from the rotational pole, then
the attachment to different field lines would produce an eastward drift as
observed.  That said, these MHD simulations appear to have strong, chaotic
dependencies on their parameters.  (For $\alpha = 15\degr$, the hot spot is
south of the magnetic pole; for 30\degr, west; and for 45\degr, north!)  More
work is needed to better model these observations.

This scenario does not explain why the shift of the pulse peak would solely be
expressed by a change in the fundamental; during these episodes in 2002 and
2005 the phase of the harmonic remains relatively constant.  However, a
movement of the hotspot toward the magnetic pole would most likely change the
shape of the hotspot, possibly in a way that would preserve the phase of the
harmonic.

The slow drifts seen during the tails of the 1998 and 2002 outbursts are also
difficult to explain.  The flares during the tail cause the luminosity to
change in a periodic manner, so we cannot expect a monotonic motion of the
accretion disk's inner edge.  The net drift during the tail of the 2002
outburst is of the same magnitude as the rapid phase shift that happens right
before the tail begins, suggesting the drift may be a relaxation of the
accretion column back to its original location.

\subsection{Comparison with previous spin\\frequency measurements}

There have been a number of previous reports of short-term $\dot\nu$
measurements made during outbursts of several accreting millisecond pulsars
including XTE~J0929$-$314 \citep{Galloway02}, \saxj\ \citep{Morgan03,
Burderi06}, XTE~J1751$-$305 \citep{Markwardt03}, IGR~J00291+5934
\citep{Falanga05, Burderi07}, and XTE~J1814$-$334 \citep{Papitto07}.  Some of
the reported $\dot\nu$ values have been surprisingly large given the estimates
of $\dot M$ during the outbursts, possibly violating a basic prediction of
magnetic disk accretion theory: that accretion torques cannot exceed the
characteristic torque $N_{\rm char}=\dot M(G M r_{\rm co})^{1/2}$ exerted by
accreting Keplerian material at the corotation radius
\citep[e.g.,][]{Ghosh79b}.

In the particular case of \saxj, spin derivatives as large as a few times
$10^{-13}$ Hz~s$^{-1}$ near the outburst peak were reported \citep{Morgan03,
Burderi06}, corresponding to accretion torques exceeding $N_{\rm char}$ for
this source.  However, these studies calculated pulse phase residuals using
only a single harmonic; \citet{Morgan03} reported $\dot\nu$ detections using
only the fundamental, while \citet{Burderi06} measuring the phase from the
second-harmonic alone after noting the sudden phase shift of the fundamental
in the middle of the 2002 outburst.  Our results in \S3.6 indicate that
both of these approaches are likely to be contaminated by pulse shape changes,
at least in the case of \saxj.

\begin{figure}[t]
  \begin{center}
    \includegraphics[width=0.47\textwidth]{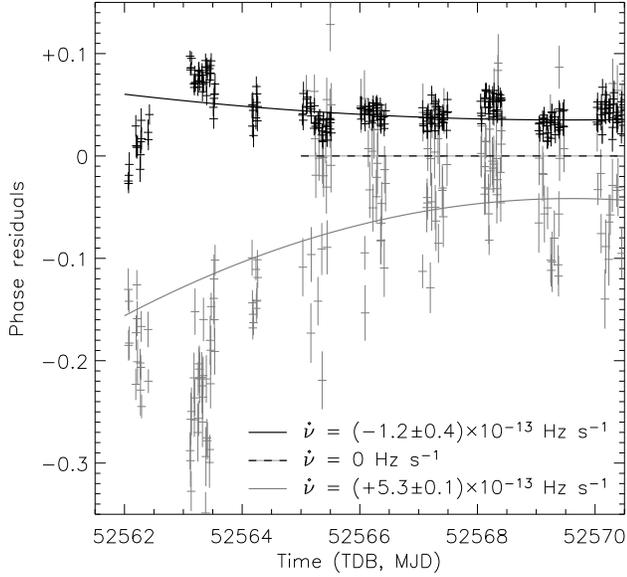}
  \end{center}
  \caption{ Fitting a frequency model using only the fundamental (black) or
    the harmonic (grey) produces non-zero $\dot\nu$ measurements during the
    peak of the 2002 outburst.  The data points are the 512~s phase residuals
    relative to the best constant-frequency model for the 2002 outburst.  The
    solid lines give the best constant-$\dot\nu$ models, fit solely to the
    fundamental or the second harmonic.  The dashed line shows the
    constant-frequency model derived using both, combined via
    equation~(\ref{eq:weighting}); this fit did not use the points prior to
    MJD 52565.
  \label{fig:badfdots}}
\end{figure}

Figure~\ref{fig:badfdots} illustrates this point.  Taking the phases of the
harmonic components as direct spin measurements can produce large values of
$\dot\nu$ during the peak of the 2002 outburst.  Fitting only the
fundamental's phase residuals during the first 10~d of the outburst, we find
$\dot\nu = (-1.2\pm0.4)\times10^{-13}$~Hz~s$^{-1}$. On the other hand, using
only the second harmonic for the same interval, we find $\dot\nu =
(5.3\pm0.1)\times10^{-13}$~Hz~s$^{-1}$, in good agreement with the
\citet{Burderi06} measurement.  Because the pulse shape is changing rapidly
during this part of the outburst, the pulse arrival times cannot be accurately
determined.  We therefore cannot reliably use this part of the outburst to
measure the spin of the NS.  Note that if we exclude this region of large
pulse shape variability, the remaining phase residuals are consistent with a
constant spin frequency over the outburst interval (\S3.6).  From an
examination of all the outbursts of \saxj\ (excluding regions of large pulse
shape variability), our work sets an upper limit of $|\dot\nu| \lesssim
2.5\times 10^{-14}$~Hz~s$^{-1}$.

We thus conclude that the past measurements of short-term $\dot\nu$ in \saxj\
are unreliable.  The analysis technique we described in \S2.3 can mitigate the
effects of pulse shape variability to some extent, but attempts to measure
$\dot\nu$ in accreting pulsars must properly account for these variability
effects, and in some instances these effects may prevent such measurements.
The $\dot\nu$ measurements reported in other accreting millisecond pulsars
must all be reevaluated in this light; all the apparent violations of the
$N\leq N_{\rm char}$ limit predicted by theory may be owing to spurious
measurements caused by pulse shape variability.  However, at least some
accreting millisecond pulsars are observed to have relatively stable pulse
shapes, indicating that accurate short-term $\dot\nu$ measurements are
possible and that previous measurement of these sources should be reliable.

\subsection{Constraints on the magnetic field}
\label{sect:BFieldConstraints}

We showed in \S4.1.1 that the condition $N_{\rm dipole}\leq N_{\rm obs}$
implies that the magnetic dipole moment $\mu \lesssim 0.8\times
10^{26}$~G~cm$^3$.  This limit is consistent with the range for $\mu$ implied
by the observation of accretion-powered pulsations throughout the outbursts
\citep{Psaltis99}.  At low accretion rates, the field cannot be so strong that
it centrifugally inhibits matter from reaching the NS; during times of high
accretion, it must be strong enough to truncate the disk above the stellar
surface in order for there to be pulsations.  The dimmest observation in which
we observed pulsations was in 1998, with a flux in the 2--25~keV band of
$1.5\times10^{-11}$\fluxunits; the brightest was at the peak of the 2002
outburst, $2.62\times10^{-9}$\fluxunits.  These fluxes, along with an improved
estimate of the Eddington luminosity from observations of photospheric radius
expansion bursts \citep{Galloway06}, give us new limits on the range of
accretion rates at which pulsations have been detected, relative to the
Eddington rate $\dot{M}_{\rm E}$:  $\dot{M}_{\rm min} = 1.8\times10^{-4}\
\dot{M}_{\rm E}$ and $\dot{M}_{\rm max} = 0.03\ \dot{M}_{\rm E}$.  (We have
made the usual assumption that $L \propto \dot{M}$.)  These limits allow us to
update the range for $\mu$ derived in \citet{Psaltis99}, equations~(11) and
(12):\footnote{In deriving this range for $\mu$, we make the same conservative
assumptions as \citet{Psaltis99}: the \citet{Ghosh91} boundary layer parameter
ranges on $0.1 < \gamma_B(\dot{M}) < 1$; the NS mass is $1.4\ M_\sun < M <
2.3\ M_\sun$; and the NS radius is $10\textrm{ km} < R < 15\textrm{ km}$.}
\begin{equation}
  0.2\times10^{26}\textrm{ G cm}^3 \lesssim \mu \lesssim
  6\times10^{26}\textrm{ G cm}^3\, .
\end{equation}
Taken together with the $N_{\rm dipole}$ limit, we obtain a fairly narrow
allowed range for the magnetic dipole moment,
\begin{equation}
  0.2\times10^{26}\textrm{ G cm}^3 \lesssim \mu \lesssim
  0.8\times10^{26}\textrm{ G cm}^3\, ,
\end{equation}
which corresponds to a surface dipole magnetic field strength of
$(0.4$--$1.5)\times 10^8$~G.  This field is relatively weak: the magnetic
fields implied by the Austrilia Telescope National Facility Pulsar
Catalog\footnote{\url{http://www.atnf.csiro.au/research/pulsar/psrcat/}\\
Pulsars associated with clusters were excluded to minimize the impact of
line-of-site accelerations.  Field strengths were approximated using
equation~(\ref{eq:spindowndipole})} \citep{ATNFPsrCat} for millisecond pulsars
range from $1.1\times10^8$~G to $14\times10^8$~G.

\subsection{Constraints on accretion torques}

Even though we did not detect an accretion-induced $\dot\nu$ during the
outbursts, our new upper limits on $|\dot\nu|$ provide far stronger
constraints on the accretion physics of low-$B$ systems such as \saxj\ than
previous measurements.  Following the earlier analysis of the 1998 outburst by
\citet{Psaltis99}, the lower limit on the spin frequency derivative predicted
by accretion torque theory during an outburst with an average accretion rate
of $\dot{M}_{\rm avg} \approx \frac{1}{3}\dot{M}_{\rm max} \approx 0.01\
\dot{M}_{\rm E}$ is
\begin{eqnarray}
  \dot\nu & \gtrsim & 2\times10^{-14} \; \eta
    \left(\frac{I}{10^{45}\ {\rm g\ cm^2}}\right)^{-1}
    \left(\frac{R}{10\ {\rm km}}\right)^{3/2}
    \nonumber\\ & & \times
    \left(\frac{M}{1.4\ M_\sun}\right)^{1/2}
    \left(\frac{\dot{M}_{\rm avg}}{0.01\ \dot{M}_{\rm E}}\right)
    \textrm{ Hz s}^{-1} \, ,
\end{eqnarray}
where $\eta$ is a dimensionless parameter encapsulating the disk-magnetosphere
interaction.  (Refer to \citealt{Ghosh79b} for a discussion of the physics
that goes into this parameter.)  $\eta$ is strongly dependent on the
magnetospheric radius.  For $r_0 \approx r_{\rm co}$, the NS will be in spin
equilibrium with the accreted matter and $\eta$ will be small.  From the
$\dot\nu$ confidence intervals in Table~\ref{tbl:spinparams}, the probability
that we would have missed detecting the resulting
$2\times10^{-14}$~Hz~s$^{-1}$ spin up is 0.15\%, suggesting that $\eta < 1$
and the source is near spin equilibrium during the outbursts.

\subsection{Discussion of the increasing $P_{\rm orb}$}
\label{sect:discussporb}

Our seven year baseline for timing analysis provides the most precise
measurements yet of the orbital period of \saxj.  We find that the orbital
period is increasing at a rate $\dot{P}_{\rm orb} =
3.5(2)\times10^{-12}$~s~s$^{-1}$.  This $\dot{P}_{\rm orb}$ lies somewhat
outside the 90\% confidence upper limit set by \citet{Papitto05} using the
1998--2002 outbursts, most likely owing to the more limited baseline available
in that analysis.

It is interesting to compare our measurement with theoretical expectations.
For orbital periods $\lesssim$3~hr, mass transfer is LMXBs is driven by
angular momentum losses due to gravitational radiation from the binary
\citep{Kraft62}, since magnetic braking torques are thought to be ineffective
in this regime \citep{Rappaport83, Spruit83}.  For \saxj, the $\dot M$
predicted by this mechanism is consistent with observationally inferred
long-term average value of $\dot{M} = 1\times10^{-11}\ M_\sun$~yr$^{-1}$
\citep{Bildsten01}.  For conservative mass transfer from a degenerate (brown
dwarf) donor, this predicts orbital expansion on a time scale $P_{\rm orb} /
\dot{P}_{\rm orb}= 3$~Gyr \citep[see, e.g.,][]{Tauris06}.  By contrast, our
measured value of $P_{\rm orb}/\dot{P}_{\rm orb} = 66$~Myr is an order of
magnitude more rapid.

The origin of the anomalously large $\dot{P}_{\rm orb}$ in \saxj\ is unclear,
although we note that unexpectedly large $\dot{P}_{\rm orb}$ values have also
been observed in several other LMXBs including 4U~1820$-$30
\citep{VanDerKlis93}, EXO~0748$-$676 \citep{Wolff02}, and 4U~1822$-$371
\citep{Hellier90}.  As pointed out by \citet{Chakrabarty98}, the binary
parameters of \saxj\ are very similar to those of the so-called ``black
widow'' millisecond radio pulsars, all of which are ablating their low-mass
companions \citep[see, e.g.,][]{Fruchter90}.  If \saxj\ does indeed turn on as
a radio pulsar during X-ray quiescence (\citealt{Burderi03};
\citealt{Campana04}; see also \S4.1.1), it may be a black widow system as
well, consistent with its very low donor mass.  As such, it is interesting to
note that a large and variable $\dot{P}_{\rm orb}$, both positive and
negative, has been measured in two black widow pulsars \citep{Arzoumanian94,
Doroshenko01}.

Although mass loss from the companion through an ablated wind would tend to
increase $\dot{P}_{\rm orb}$, the mass loss rate required to explain the
observed $\dot{P}_{\rm orb}$ in \saxj\ is $\sim$$10^{-8} M_\odot$~yr$^{-1}$
\citep{Tauris06}; this is unphysically large given our measured pulsar
spindown rate (\S4.1), which sets the pulsar luminosity available for
irradiating the companion. This explanation for $\dot{P}_{\rm orb}$ is also
inadequate in the black widow pulsars, where the orbital period variability is
quasi-cyclic on a $\simeq$10~yr time scale \citep{Arzoumanian94,
Doroshenko01}.  In those systems, it has been suggested that tidal dissipation
and magnetic activity in the companion is responsible for the orbital
variability, requiring that the companion is at least partially
non-degenerate, convective, and magnetically active \citep{Arzoumanian94,
Applegate94, Doroshenko01}.  If this mechanism is active in \saxj, we would
expect quasi-cyclic variability of $P_{\rm orb}$ to to reveal itself over the
next few years.

\bigskip
\acknowledgements{We thank Lars Bildsten, Andrew Cumming, Andrew King, Miriam
Krauss, Fred Lamb, Juri Poutanen, Dimitrios Psaltis, Anatoly Spitkovsky, and
Anna Watts for useful discussions.  This work was supported in part by NASA.}

\begin{appendix}
\section{Improved Optical Position for \saxj} \label{app:newpos}

An accurate source position is essential for high-precision pulsar timing.  An
incorrect position results in errors during the barycentering of X-ray arrival
times, producing frequency offsets due to improperly corrected Doppler shifts
\citep[see, e.g.,][]{Manchester72}.  \saxj\ lies only $\beta = -13.6^\circ$
below the ecliptic plane, so any errors during barycentering will be
particularly pronounced.  For example, a position error of $\epsilon =
0\farcs2$ parallel to the plane of the ecliptic produces frequency and
frequency derivative offsets relative to $\nu_0 \approx 401$~Hz of
\begin{eqnarray}
  \Delta\nu &=& \nu_0 \epsilon \left(a_\earth \cos \beta / c\right)
                \left(2\pi / P_\earth\right) \cos \tau
             =  40 \, \cos\tau \ \textrm{nHz} \label{eq:freqvsposition} \\
  \Delta\dot\nu &=& -\nu_0 \epsilon \left(a_\earth \cos \beta / c\right)
                \left(2\pi / P_\earth\right)^2 \sin \tau
             =  -8\times10^{-15} \sin\tau \ \textrm{Hz s}^{-1} \, .
\end{eqnarray}
Here $\tau = 2\pi t / P_\earth$ parametrizes the Earth's orbit, with time $t$
equal to zero when the Earth is closest to the source.  These offsets are
comparable with the expected timing uncertainties.  Each outburst gives a
baseline of about $2\times10^6$~s over which we can typically measure pulse
arrival times with an accuracy of better than 25~\us, or
$1\times10^{-2}$~cycles, producing $\sim$5~nHz frequency uncertainties.  By
similar logic, we should be sensitive to $\dot\nu$'s as small as
$\sim$$3\times10^{-15}$~Hz~s$^{-1}$.  In practice, the pulse shape noise
observed in \saxj\ makes the actual uncertainties somewhat greater than these
back-of-the-envelope values, increasing the $\nu$ uncertainty by a factor of
$\sim$2 and the $\dot\nu$ uncertainty by a factor of $\sim$10, but the
frequency uncertainty is still substantially less than the offsets due to a
$0\farcs2$ position error.

\begin{figure}[t]
  \begin{center}
    \includegraphics[width=0.47\textwidth]{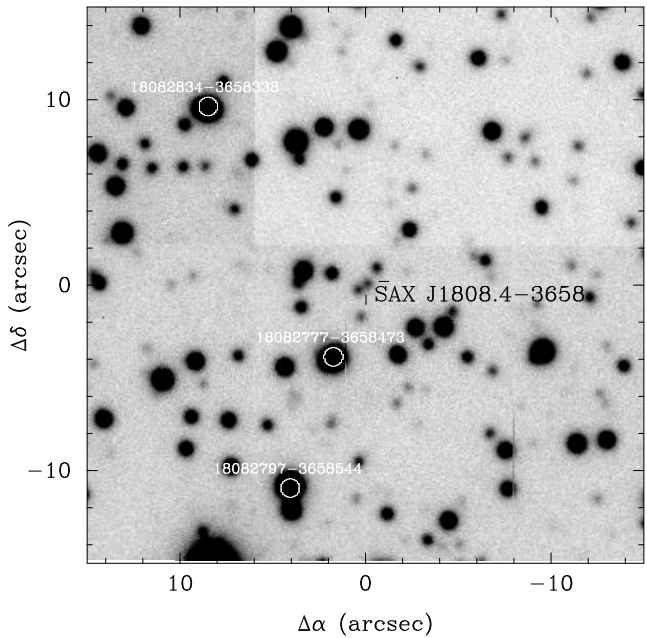}
  \end{center}
  \caption{ A $30\arcsec$ portion of our $r^{\prime}$-band Magellan image.
    The counterpart of \saxj\ is indicated by the tick marks: it is the
    north-west object of the close pair at the center.  We also indicate three
    2MASS stars that we used for astrometry with the circles.  The changing
    grayscale levels across the image reflects poor correction for the
    four-amplifier readout of MagIC but does not affect our astrometry.
  \label{fig:finder}}
\end{figure}

We observed the field of \saxj\ with the Raymond and Beverly Sackler Magellan
Instant Camera (MagIC) on the 6.5-m Baade (Magellan I) telescope on the night
of 2001~June~13, using the $r^{\prime}$ filter.  The seeing was $0\farcs5$.
Figure~\ref{fig:finder} shows the results.  After standard reduction,
involving bias-subtraction and flatfielding, we attempted to register the
field to the International Coordinate Reference System (ICRS)\@.  We examined
three astrometric catalogs for this purpose: the \textit{Hubble Space
Telescope} Guide Star Catalog (GSC, which was used by \citealt{Giles99};
\citealt{Lasker90}), the USNO-B1.0 survey \citep{Monet03}, and the Two-Micron
All-Sky Survey (2MASS; \citealt{2MASS})\@.  We selected stars from all three
catalogs that were not saturated or blended on our image, and fit using the
\texttt{IRAF} task \texttt{ccmap} for the position offset, rotation, and
plate-scale.  We found that we could obtain the best astrometry with 2MASS:
with USNO and GSC, many stars that had consistent positions between 2MASS and
our image had deviations of more than $0\farcs2$, the overall scatter was
larger, and there were fewer stars.  With 2MASS we fit using 70 stars across
the $2\arcmin$ MagIC frame.  With position residuals of $0\farcs08$ in each
coordinate, we obtained a combined uncertainty of $0\farcs08 / \sqrt{70} =
0\farcs01$.  Therefore, our astrometric uncertainty is dominated by the
$\approx$$0\farcs15$ position uncertainty of 2MASS\@.

To verify our position, we checked for stars on the MagIC image from the
Second US Naval Observatory CCD Astrograph Catalog (UCAC2;
\citealt{Zacharias04})\@.  These are highly accurate positions (individual
uncertainties of 20--40~mas) for relatively bright ($\approx$15~mag) stars
taken with a CCD at a current epoch (1996--1998) and with proper motions.  We
found three unsaturated UCAC2 stars on our image:  16259696, 16259777, and
16259680.  We measured their positions on our image and compared the positions
derived from the 2MASS solution to those from UCAC2, updated to epoch 2001.45.
We found no net shift, and the offsets are less than $0\farcs16$ in all cases.
(We note that the stars are toward the edge of the image, where residual image
distortions may be present, in contrast to \saxj\ which is at the center of
the image)\@.  Therefore we believe that our solution using 2MASS is indeed
accurate to our stated uncertainty of $0\farcs15$.

We then measured the position of \saxj\ on the image and transformed the
position to the ICRS.  The position that we find is: R.A.~= $18^{\rm h}08^{\rm
m}27\fs62$, Decl.~= $-36\degr58\arcmin43\farcs3$, equinox J2000.0, with
uncertainty $0\farcs15$.  This is $1\farcs5$ from the \citet{Giles99}
position, twice their quoted $0\farcs8$ uncertainty.  But with many more
reference stars of higher quality over a smaller field (\citealt{Giles99} used
5 GSC stars over a $4\arcmin$ field), and CCD data taken at a more recent
epoch (1998 for 2MASS, vs.\ 1987--1996 for GSC\footnote{The USNO does not
recommend GSC for current use: see
\url{http://ad.usno.navy.mil/star/star\_cats\_rec.shtml\#gsc2.2}.}  and 1981
for USNO), this new position should be more accurate.

\section{Derivation of Phase Uncertainties} \label{app:phaseerrs}

Derivation of the uncertainties of the phase residuals, as given in
equation~(\ref{eq:sigmak}), follows from our definition of the phases,
\begin{equation}
  A_k \exp\left(2\pi i k \phi_k\right) =
    2 \sum_{j=1}^{n} x_j \exp\left(2\pi i j k / n \right) \, ,
  \label{eq:phaseerr1}
\end{equation}
where we have divided our phases into $n$ bins, each containing $x_j$ photons.
Inverting to solve for $\phi_k$,
\begin{equation}
  \phi_k = \frac{1}{2\pi i k}\left(
           \ln\sum_{j=1}^{n} x_j E_{jk} - \ln\frac{A_k}{2}\right) \, ,
\end{equation}
where we define the constants $E_{jk} \equiv \exp\left(2\pi i j k / n
\right)$ for the sake of brevity.

For relatively low fractional amplitudes (certainly the case throughout this
paper), each phase bin will contain approximately the same number of photons:
$x_j \approx N_{\rm ph} / n$, with variances $(\sigma x_j)^2 \approx N_{\rm
ph} / n$ due to Poisson counting statistics.  These add in quadrature to give
the variance in $\phi_k$:
\begin{equation}
  \sigma_k^2 = \left| \sum_{j=1}^n
               \left(\frac{\partial\phi_k}{\partial x_j}\right)^2
               \left(\sigma x_j\right)^2 \right|
             = \left| \sum_{j=1}^n \left(\frac{1}{2\pi k}
	       \frac{E_{jk}}{\sum_{j'=1}^n x_{j'} E_{j'k}} \right)^2
               \left(\frac{N_{\rm ph}}{n}\right)\right| \, .
\end{equation}
Summing the exponentials, we have $\left|\sum_{j=1}^n E_{jk}^2\right| =
\frac{1}{2}n$.  From the definition of $A_k$ in equation~(\ref{eq:phaseerr1}),
$\left|\sum_{j=1}^n x_j E_{jk}\right| = \frac{1}{2} A_k$.  Substituting these
in, we reach our estimate of the phase uncertainty: $\sigma_k = \sqrt{2 N_{\rm
ph}} / 2\pi k A_k$.

\end{appendix}

\bibliography{Paper} \bibliographystyle{apj}

\end{document}